\documentclass[a4paper,fleqn,usenatbib]{mnras}

\usepackage[T1]{fontenc}
\usepackage{ae,aecompl}

\usepackage{graphicx}	
\usepackage{amsmath}	
\usepackage{amssymb}	
\usepackage[outdir=./]{epstopdf}
\usepackage{lscape}
\usepackage{rotating}

\newcommand{\zrms}{z_{\rm rms}}

\title[Bar Formation in Cosmology]{Can Stellar Discs in a Cosmological Setting Avoid Forming Strong Bars?}

\author[Jacob S. Bauer et al.]{
Jacob S. Bauer,$^{1}$\thanks{E-mail: jacob.bauer@queensu.ca}
Lawrence M. Widrow,$^{1}$
\\
$^{1}$Department of Physics, Engineering Physics \& Astronomy, Queen's University, Stirling Hall, Kingston, ON K7L 3N6, Canada
}

\date{Accepted XXX. Received YYY; in original form ZZZ}

\pubyear{2018}

\begin{document}

\label{firstpage}
\pagerange{\pageref{firstpage}--\pageref{lastpage}}
\maketitle

\begin{abstract}

We investigate the connection between the vertical structure of
stellar discs and the formation of bars using high-resolution
simulations of galaxies in isolation and in the cosmological context.
In particular, we simulate a suite of isolated galaxy models that have
the same Toomre $Q$ parameter and swing amplification parameter but
that differ in the vertical scale height and velocity dispersion.  We
find that the onset of bar formation occurs more slowly in models with
thicker discs.  Moreover, thicker discs and also discs evolved in
simulations with larger force softening also appear to be more
resilient to buckling, which acts to regulate the length and strength
of bars.  We also simulate disc-halo systems in the cosmological
environment using a disc-insertion technique developed in a previous
paper.  In this case, bar formation is driven by the stochastic
effects of a triaxial halo and subhalo-disc interactions and the
initial growth of bars appears to be relatively insensitive to the
thickness of the disc.  On the other hand, thin discs in cosmological
halos do appear to be more susceptible to buckling than thick ones and
therefore bar strength correlates with disc thickness as in the
isolated case.  More to the point, one can form discs in cosmological
simulations with relatively weak bars or no bars at all provided the
discs as thin as the discs we observe and the softening length is
smaller than the disc scale height.

\end{abstract}


\section{INTRODUCTION}

The problem of bar formation in disc galaxies tests our understanding
of cosmological structure formation and galactic dynamics.  In
principle, theories of galaxy formation should yield predictions for
the fractional distribution of bars in terms of their strength,
length, and pattern speed.  While it is often difficult to make
precise, quantitative statements about bars from observations, general
properties of their distribution have emerged (See
\citet{sellwood1993}, \citet{Sellwood2013} and \citet{BT} for
reviews).  Roughly 30-40 per cent of disc galaxies exhibit strong
bars, that is bars that dominate the disc luminosity.  Another 20 per
cent or more have relatively weak bars.  The bar fraction appears to
increase with time.  Approximately one tenth of disc galaxies between
$0.5 \le z \le 2$ have visually identifiable strong bars, which is a
factor of 3-4 smaller than the fraction in the local Universe (see
\citet{simmons2014} and references therein).  The bar fraction also
varies with galaxy type.  \citet{masters2010} find that $70\pm 5$ per
cent of the so-called passive red spirals have bars as compared to a
$25\pm 5$ per cent bar fraction for blue spirals.  Since the red
spirals are interpreted as old galaxies that have used up their
star-forming gas, this result is consistent with a bar fraction that
increases with time.  The upshot of these observations is that in
terms of bars, disc galaxies in the local Universe divide into three
roughly equal bins: those with strong bars, those with weak bars, and
those with no detectable bar.  These observations suggest that bars are 
capable of forming at a wide range of cosmic times, but once formed,
are difficult to destroy.

Intuition tells us that properties of a bar should depend on the
properties of its host galaxy and the environment in which that
galaxy lives.  Theoretical arguments indicate that cold, thin discs
are susceptible to local ``Toomre'' instabilities.  Furthermore, discs
that are strongly self-gravitating, that is discs that contribute the
bulk of the gravitational force required to maintain their rotational
motion, are susceptible to global instabilities.  Thus, one can
construct initially axisymmetric galaxy models that form bars with
vastly different properties (or no bars at all) by changing the
internal disc dynamics or trading off disc mass for bulge and halo
mass.  The implication is that the distribution of bars provides an
indirect means for inferring a disc's kinematics and mass-to-light
ratio as well as the distribution of mass in a galaxy's dynamically
``hot'' components, namely its bulge and dark matter halo.

A galaxy's ability to resist local instabilities is typically
expressed in terms of the (kinetic) Toomre $Q$-parameter
\citep{ToomreParameter} while its ability to resist global
instabilities is encapsulated in the swing-amplification $X$-parameter
\citep{GoldreichTremaine1978,GoldreichTremaine1979}.  Both parameters
are defined so that large values imply a more stable disc.  The
hypothesis that they determine a galaxy's susceptibility to bar
formation has been tested by simulations of isolated, idealized galaxy
models \citep{PeeblesOstriker1973, ZangHohlBars1978,
  CombesSandersBars1981,Sellwood1981}.  Typically, the initial galaxy
is represented by an N-body (Monte Carlo) realization of an equilibrium solution to
the collisionless Boltzmann equation comprising a disc, dark matter
halo, and often, a bulge.  Equilibrium does not imply
stability, and a galaxy can develop spiral structure and a bar through
instabilities that are seeded by shot noise
\citep{EfstathiouShotNoise} and amplified by feedback loops such as
swing amplification \citep{Sellwood2013}. A common way to suppress the mechanisms
which give rise to these effects is by increasing either $Q$ or $X$.  For
example, in dynamically warm discs (that is, discs with high $Q$)
perturbations are randomized on timescales short enough to prevent the
feedback loop from starting \citep{AthanassoulaSellwood1986}.
Likewise, submaximal discs, that is, discs with high $X$, avoid the
bar instability presumably because the disc lacks the self-gravity to
drive the bar instability \citep{EfstathiouShotNoise,
  ChristodoulouStability1995, Sellwood2013}.

As one might imagine, the parameters $Q$ and $X$ do not uniquely
describe a galaxy's susceptibility to bar formation.
\citet{WPDGalactICSReference} present a grid of models in the $Q-X$
plane that all satisfy observational constraints for the Milky Way.
These simulations confirm the basic notion that susceptibility to
instabilities increases with decreasing $Q$ and $X$.  However, a
careful study of bar strength and length as a function of time across
these simulations suggests a more complicated picture.  In particular,
the bar strength is not a perfectly monotonic function of $X$ at fixed
$Q$ or vice versa.  The implication is that additional parameters are
required to fully predict how bar formation will proceed from some
prescribed initial conditions.  In short, bar formation may
proceed very differently within a family of models that have the same
$Q$ and $X$ but vary in other ways.

One property of a disc not captured by either $Q$ or $X$ is its
thickness, or alternatively, its vertical velocity dispersion.  (The
Toomre parameter depends only on the radial velocity dispersion.)
\citet{Klypin2009} use a suite of simulations to demonstrate that the
thickness of the disc plays a profound role in the development of a
bar.  In particular, their thick disc model forms a stronger and more
slowly rotating bar as compared with the bar that forms in a thin disc
model with the same initial radial dispersion profile and rotation
curve decomposition.  Moreover, simulation parameters such as mass
resolution and time step also influence the growth of the bar
instability and slowdown of the bar due to angular momentum transfer
with the dark halo \citep{dbs2009}.

Of course, galaxies are neither isolated nor born as axisymmetric,
equilibrium systems. In these idealized systems, instabilities can only come from shot noise, whereas this is is not true in a complicated dynamical environment. As such, bar formation may be very different in idealized
galaxies as compared with galaxies in a cosmological setting.  For
example, halo substructure in the form of satellite galaxies and dark
matter subhaloes can pass through and perturb the disc.
\citet{gauthier2006}, \citet{kazantzidis2008}, and
\citet{dbs2009} showed that an apparently stable disc galaxy
model can develop a bar when a fraction of the ``smooth'' halo is
replaced by substructure in the form of subhaloes.  The effect is
stochastic; subhalo-triggered bar formation seems to require subhaloes
whose orbits take them into the central regions of the disc in a
prograde sense.  More recently, \citet{purcell2011} showed that
Sagittarius dwarf alone could have been responsible for the Milky
Way's spiral structure and bar.

Cosmological haloes also possess large-scale time-dependent tidal
fields, which impart torques to the disc and cause it to precess,
nutate, and warp
\citep{dubinski1995,binney1998,dubinski2009,Bauer2018a}.  In turn,
stellar discs can reshape the inner parts of the dark matter haloes
via adiabatic contraction and dynamical friction
\citep{blumenthal1986, ryden1987, dubinski1994,
  DubinskiKuijkenRigidDisks,
  DeBuhrStellarDisks,YurinSpringelStellarDisks, Bauer2018a}.  In
principle, one can turn to {\it ab initio} hydrodynamic cosmological
simulations to capture the effects of both triaxiality and
substructure.  Indeed, galaxy formation studies in the cosmological
environment that include the formation of supermassive black holes,
stars, and the feedback from these objects on galaxy formation have
had remarkable success over the last decade (See, for
  example \citet{IllustrisFeedback, Eagle}).  Unfortunately, feedback
techniques are extremely computationally expensive.  Moreover, the
simulator cannot control the properties of the disc such as its mass
and radial scale length that form within a particular haloes.  This
restriction makes it difficult to explore the ``nature vs. nurture''
question: Do bars reflect the structure of their host galaxy,
substructure interactions of the disc's lifetime, or large-scale tidal
fields from the dark halo?

Techniques developed by \citet{BerentzenShlosmanStellarDisks},
\citet{DeBuhrStellarDisks}, \citet{YurinSpringelStellarDisks},
\citet{Bauer2018a} and others allow one to insert a collisionless disc
into a dark matter halo.  These techniques provide a compromise
between fully cosmological simulations and simulations of isolated
galaxies.  In general, the first step is to run a pure dark matter
simulation of a cosmological volume and select a halo suitable for the
galaxy one intends to study.  The simulation is then rerun with a rigid
disc potential that is adiabatically grown from some early epoch (say
redshift $z=3$) and an intermediate one (say $z=1$).  Doing so allows
the halo to respond to ``disc formation''.  At the intermediate
redshift, a suitable N-body disc that is approximately in equilibrium
is swapped in for the rigid disc potential and the simulation
continues, now with live disc and halo particles.

Perhaps the most striking and perplexing result to come from recent
disc-insertion simulations is the prevalence of strong bars.
\citet{YurinSpringelStellarDisks}, for example, find that all of the
discs in their bulge-less simulations for strong bars even though some
of these discs are decidedly submaximal.  In addition, over half of
the discs in simulations with classical bulges form strong bars.
Their results suggest that discs in the cosmological setting are more
prone to forming strong bars and that bulges play an essential role in
explaining the presence of disc galaxies with weak bars or no bars at
all.

Though the \citet{YurinSpringelStellarDisks} models vary in $Q$ and
$X$ they share two important properties.  First, the ratio of the
radial and vertical velocity dispersion is set to unity throughout the
disc.  Second, the ratio of the vertical and radial scale lengths is
fixed at $0.2$, which is roughly a factor of two larger than that of
the Milky Way's thin disc.  In addition, the softening length in their
simulations is fixed at $680\,{\rm pc}$.  Thus, the discs in their
simulations are thicker and (vertically) warmer than what one might
expect for Milky Way-like galaxies.  The results from
\citet{Klypin2009} suggest that these properties rather than (or
together with) halo substructure and triaxiality are responsible for
the preponderance of strong bars in the
\citet{YurinSpringelStellarDisks} models.

In this paper, we attempt to isolate the different effects that
determine whether a galaxy forms a strong bar, a weak one, or no bar
at all.  The core of the paper is a a sequence of N-body simulations
that include simulations of isolating disc galaxies and galaxies in
cosmological haloes.  Our choice of models is meant to complement
those of \citet{YurinSpringelStellarDisks}.  In particular, we choose
models that have the same $Q$ and $X$ but that vary in 
their vertical structure.  

We begin in section \S \ref{sec:ics} by discussing the dimensionless
parameters that arise when one constructs equilibrium disc galaxy
models.  These include the aforementioned $Q$ and $X$ parameters as
well as the ratio of the vertical and radial velocity dispersion and
the ratio of the vertical and radial scale lengths.  We also discuss
possible effects of gravitational softening.  In
\ref{sec:thick_discs_suppress} we describe our sequence of simulations
and discuss how they fit in with previous work.  We discuss results
for our isolated galaxy simulations in \S \ref{sec:isolated} and our
cosmological ones in \S \ref{sec:cosmo}.  We conclude with a
discussion of the implications of our results in \S
\ref{sec:conclusions}.

\begin{figure}
	\includegraphics[width=0.5\textwidth]{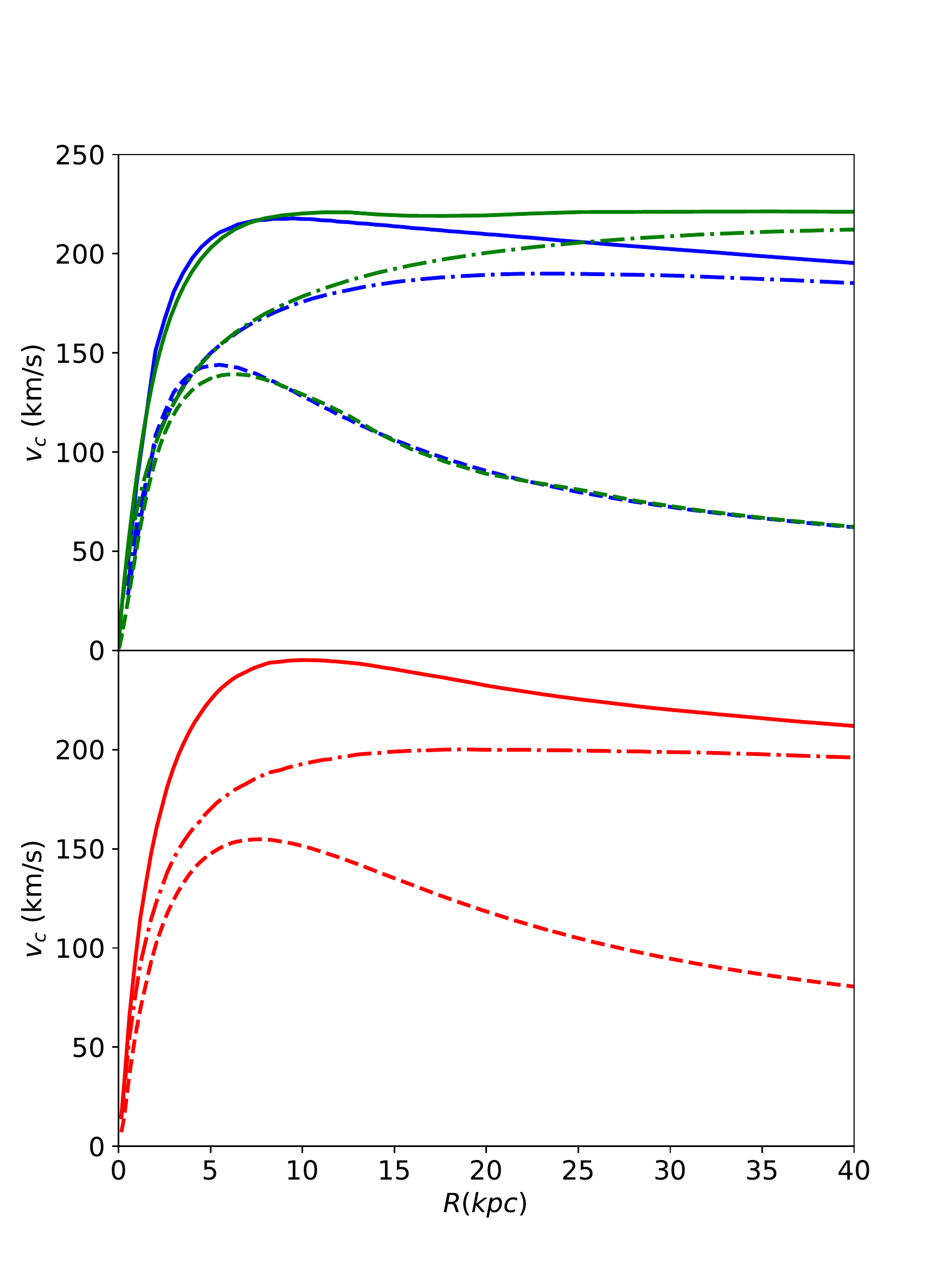}
	\caption{Rotation curve decomposition for our models. Total
          rotation curves are shown as solid lines while the separate
          contributions from the disc and halo are shown as dashed and
          dot-dashed curves, respectively.  Blue curves in the top
          panel are for the isolated galaxy simulations with
          \textsc{GalactICS} initial conditions while the green curves
          are for the simulations C.I.Ag run with \textsc{AGAMA}
          initial conditions.  Bottom panel shows initial rotation
          curve decomposition for the runs D.I and
          E.II.}\label{fig:rcs}
\end{figure}

\begin{table*}
\begin{tabular}{l l l l l l l l l l l l}
\hline
 & $M_d$ & $R_d$ & $V_c$ & $\sigma_R$ & $z_d/R_d$  & $\sigma_R/\sigma_z$  & $X$ &  $Q$ & $\rho_h/\rho_0$ & $\epsilon$ \\ 
\hline
A.I     & 3.49 & 2.50 & 216 & 25.3 & 0.10 & 1.27 & 2.34 & 1.00 & 0.14 & 0.15\\
A.II    & 3.49 & 2.50 & 216 & 25.3 & 0.10 & 1.27 & 2.34 & 1.00 & 0.14 & 0.50\\
B.I     & 3.49 & 2.50 & 213 & 25.3 & 0.20 & 0.97 & 2.34 & 1.00 & 0.28 & 0.15\\
B.II    & 3.49 & 2.50 & 213 & 25.3 & 0.20 & 0.97 & 2.34 & 1.00 & 0.28 & 0.50\\
C.I     & 3.49 & 2.50 & 208 & 25.3 & 0.40 & 0.77 & 2.34 & 1.00 & 0.50 & 0.15\\
C.I.Ag  & 3.49 & 2.50 & 216 & 25.3 & 0.40 & 0.77 & 2.34 & 1.00 & 0.50 & 0.15\\
D.IV    & 5.82 & 3.70 & 245 & 25.2 & 0.10 & 1.14 & 2.45 & 1.00 & 0.29 & 0.18\\
E.IV    & 5.82 & 3.70 & 245 & 27.4 & 0.25 & 0.72 & 2.45 & 1.00 & 0.73 & 0.74\\
\hline
YS15.A5 & 5.00 & 3.00 & 263 & 30.7  & 0.2 & 1.00 & 3.22 & 1.38 & 0.21 & 0.68\\
YS15.B5 & 5.00 & 3.00 & 211 & 26.6  & 0.2 & 1.00 & 2.06 & 0.96 & 0.11 & 0.68\\
YS15.C5 & 5.00 & 3.00 & 270 & 30.3  & 0.2 & 1.00 & 3.31 & 1.42 & 0.23 & 0.68\\
YS15.D5 & 5.00 & 3.00 & 236 & 26.6  & 0.2 & 1.00 & 2.58 & 1.12 & 0.16 & 0.68\\
YS15.E5 & 5.00 & 3.00 & 233 & 27.1  & 0.2 & 1.00 & 2.58 & 1.11 & 0.15 & 0.68\\
YS15.F5 & 5.00 & 3.00 & 219 & 27.0  & 0.2 & 1.00 & 2.22 & 1.02 & 0.11 & 0.68\\
YS15.G5 & 5.00 & 3.00 & 227 & 28.2  & 0.2 & 1.00 & 2.45 & 1.09 & 0.13 & 0.68\\
YS15.H5 & 5.00 & 3.00 & 244 & 28.6  & 0.2 & 1.00 & 2.85 & 1.21 & 0.16 & 0.68\\
\hline
G06     & 7.77 & 5.57 & 226 & 17.1  & 0.06  & 1.80 & 2.78 & 1.43 & 0.10 & 0.15 \\
\hline
\end{tabular}
\caption{Summary of parameters for the simulations considered in this
  paper, the disk-halo simulations considered in
  \citet{YurinSpringelStellarDisks} (labeled YS15) and the
  \citet{gauthier2006} (G06).  $M_d$ is the final disc mass in units of
  $10^{10}\,M_\odot$ , $R_d$ is the disc scale radius in units of
  ${\rm kpc}$, and $V_c$ and $\sigma_R$ are the circular speed and
  radial velocity dispersion in units of ${\rm km\,s^{-1}}$ and
  evaluated at $R_p = 2.2R_d$.  For the disc aspect ratio, we quote
  $z_d/R_d$ where $z_d$ is the ${\rm sech}^2$-scale length.  The
  velocity dispersion ratio $\sigma_R/\sigma_z$, the $X$ and $Q$
  parameters, the ratio of the halo density in the midplane to that of
  the disc, and the logarithmic derivative of the circular speed are
  also measured at $R_p$.  Finally, the softening length $\epsilon$ is
  given in units of ${\rm kpc}$.  Simulations A.III and B.III are the
  same as A.I and B.I except that they are run with vertical motions
  isotropized so as to shut off the buckling
  instability.} \label{tab:sims}
\end{table*}

\section{THEORETICAL CONSIDERATIONS}
 \label{sec:ics}

In this section, we consider the structural properties of disc-halo
models with an eye toward understanding the formation of bars in these
systems.  We begin with the $Q$ and $X$ parameters and then discuss the
vertical structure of the disc as defined by its scale height,
vertical velocity dispersion, and surface density.  Finally, we
consider the effect softening might have on bar formation.

\begin{figure}
        \includegraphics[width=0.5\textwidth]{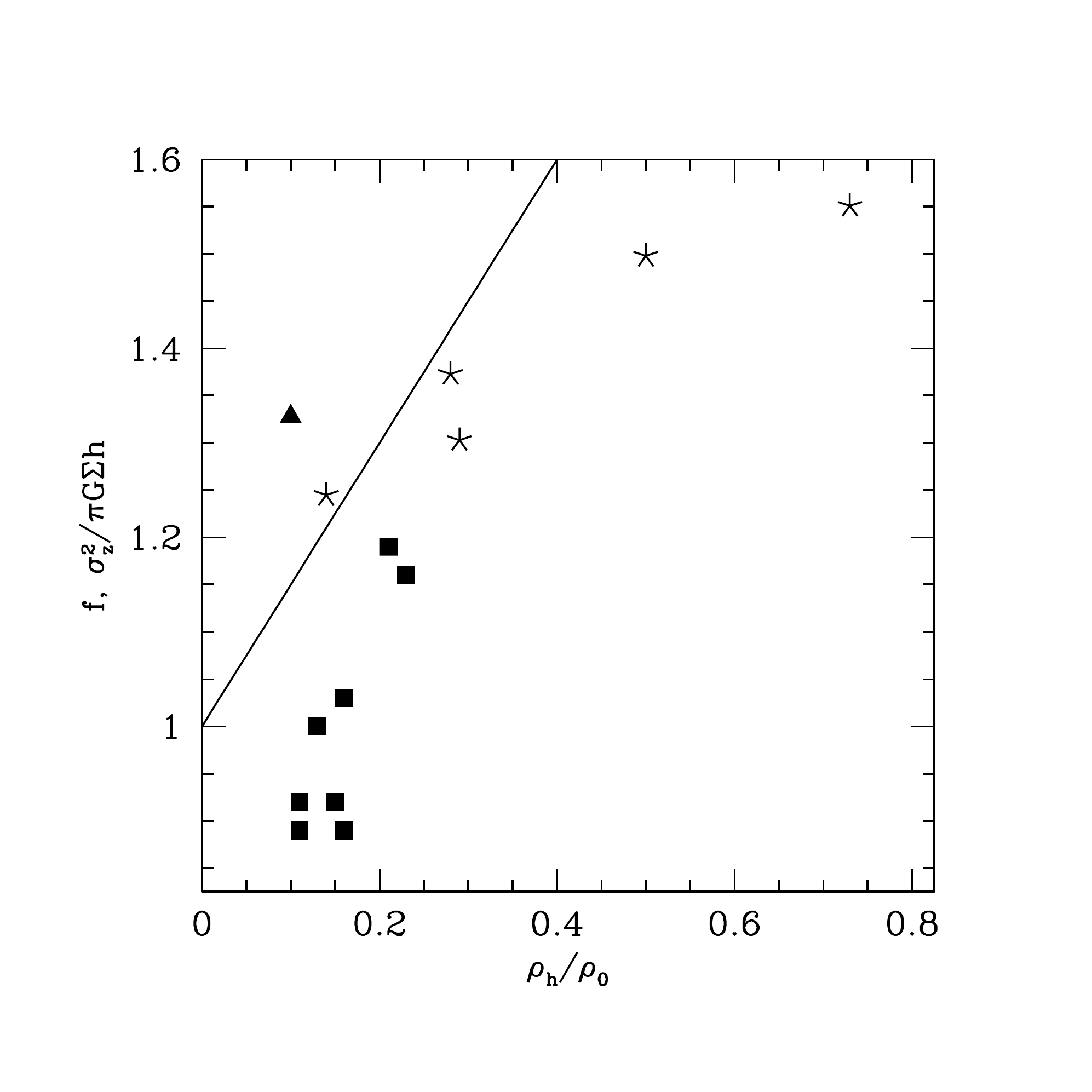}
        \caption{The dimensionless ratio $\sigma_z^2/\pi G\Sigma z_d$ as a
          function of $\rho_h/\rho_0$ for the models considered in
          this paper (stars), the disc-halo models from
          \citet{YurinSpringelStellarDisks} (filled squares) and the
          model from \citet{gauthier2006} (filled triangle).  The
          straight line is the function $f = 1 + \left (2\pi/3\right
          )^{1/2}\rho_h/\rho_0$ discussed in the text.}
\label{fig:falpha}\end{figure}

\begin{figure*}
        \includegraphics[width=1.\textwidth]{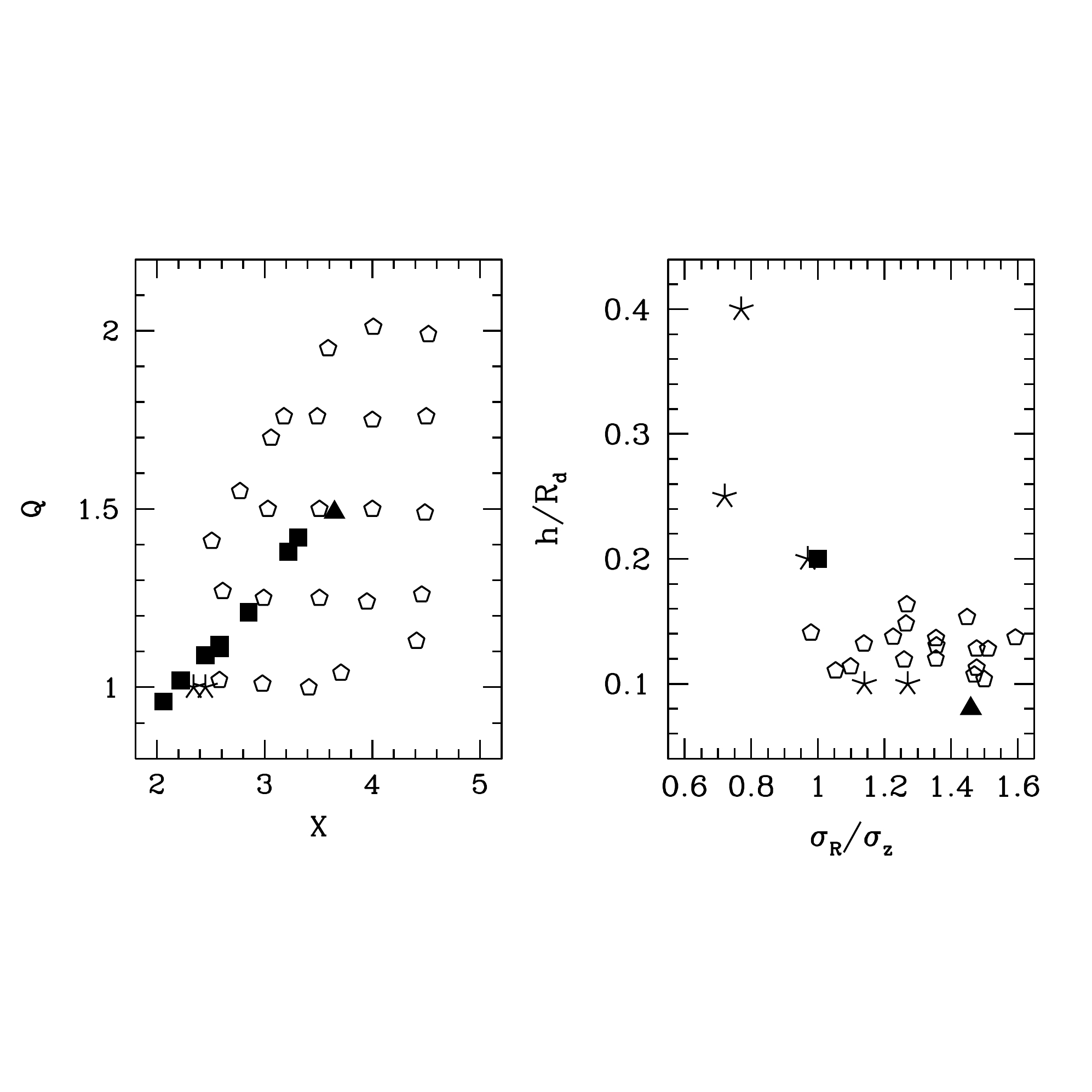}
        \caption{Distribution of simulations considered in this paper
          in the $Q-X$ and the $z_d/R_d-\sigma_R/\sigma_z$ planes.
          Stars are simulations run for this paper (A-E); filled
          squares denote the series of simulations described in
          \citet{YurinSpringelStellarDisks}; the filled triangle denotes
          the simulation of M31 run in \citet{gauthier2006}; open
          pentagons denote the simulations described in
          \citet{WPDGalactICSReference}.
        }\label{fig:QXR}\end{figure*}

\begin{figure}
	\includegraphics[width=0.5\textwidth]{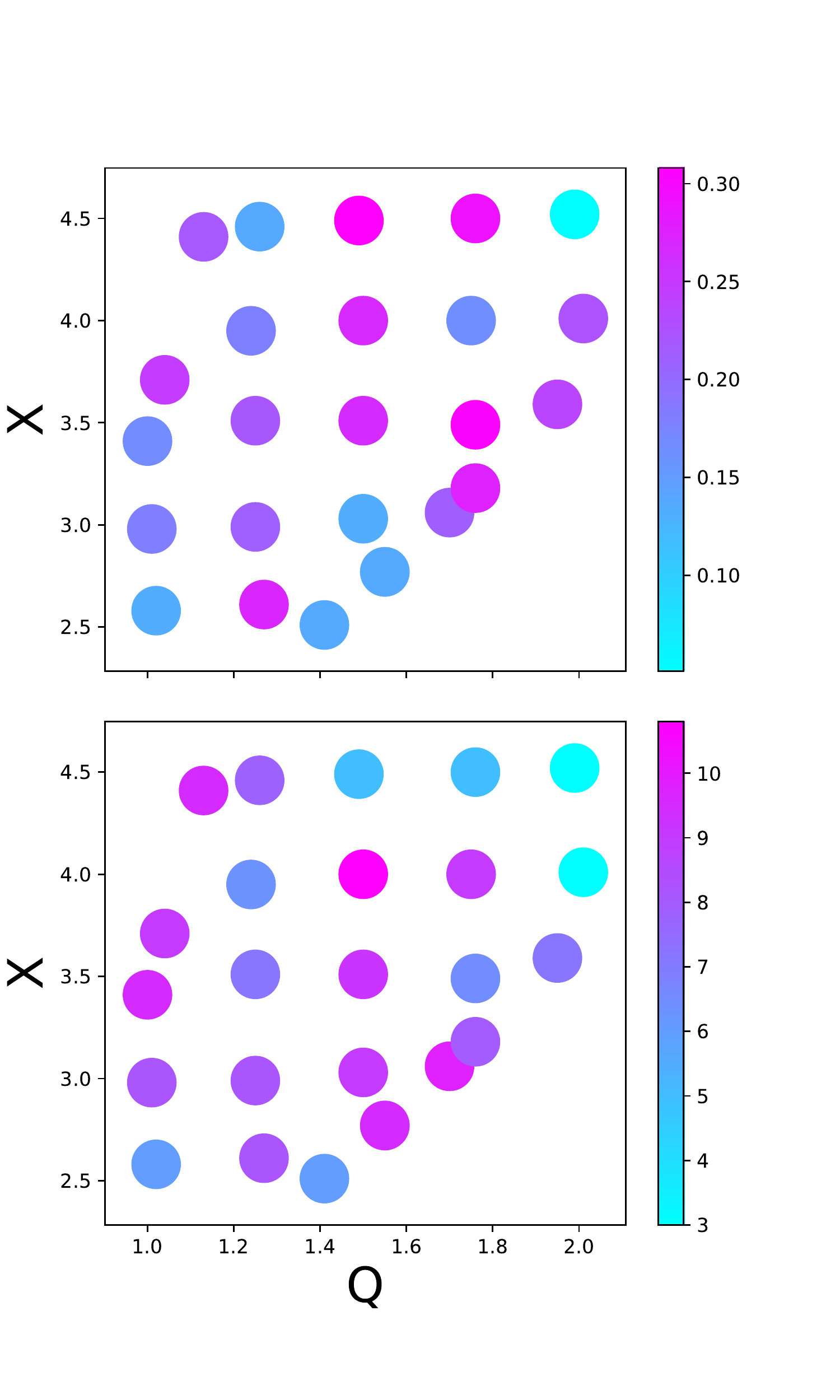}
	\caption{Strength and length of bars for the simulations
          considered in \citet{gauthier2006}.  The twenty-five models
          span the $Q$-$X$ plane.  Top panel shows the $A_2$ parameter
          while the bottom panel shows the bar length.  Both are
          measured at $5\,{\rm Gyr}$ (the final snapshot of the
          simulations).}
	\label{fig:qxa2}
\end{figure}

\begin{figure*}
	\includegraphics[width=\textwidth]{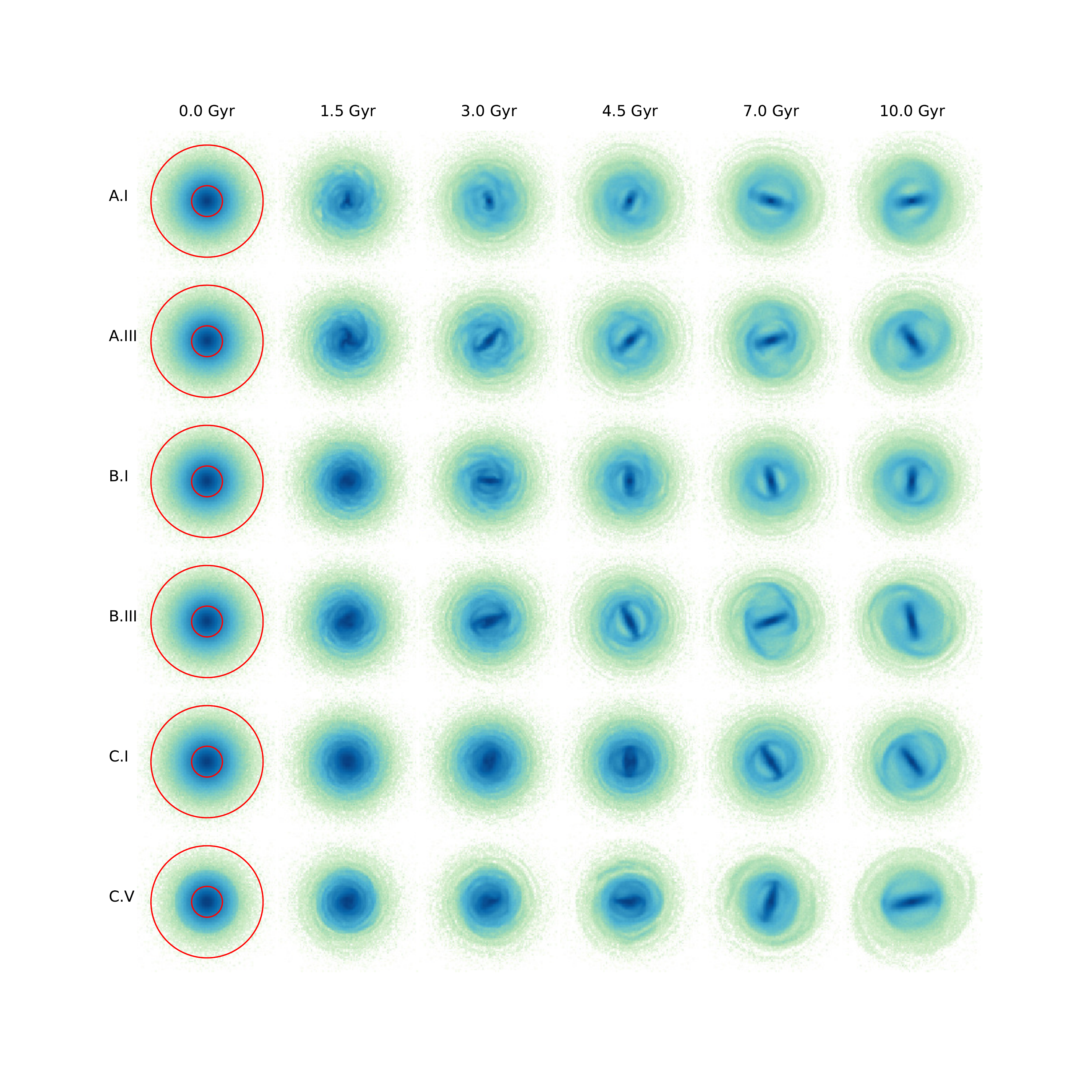}
	\caption{Surface density maps for isolated galaxy simulations
          at select times. Time proceeds from 0 to 10 Gyr,
          left-to-right, and the models span top-to-bottom in order of
          their appearance in Table \ref{tab:sims}. The overlaid red
          circles have radii $R_p=5.5\,{\rm kpc}$ and 20 kpc.}
	\label{fig:face_on_isolated}
\end{figure*}

\begin{figure}
	\includegraphics[width=0.5\textwidth]{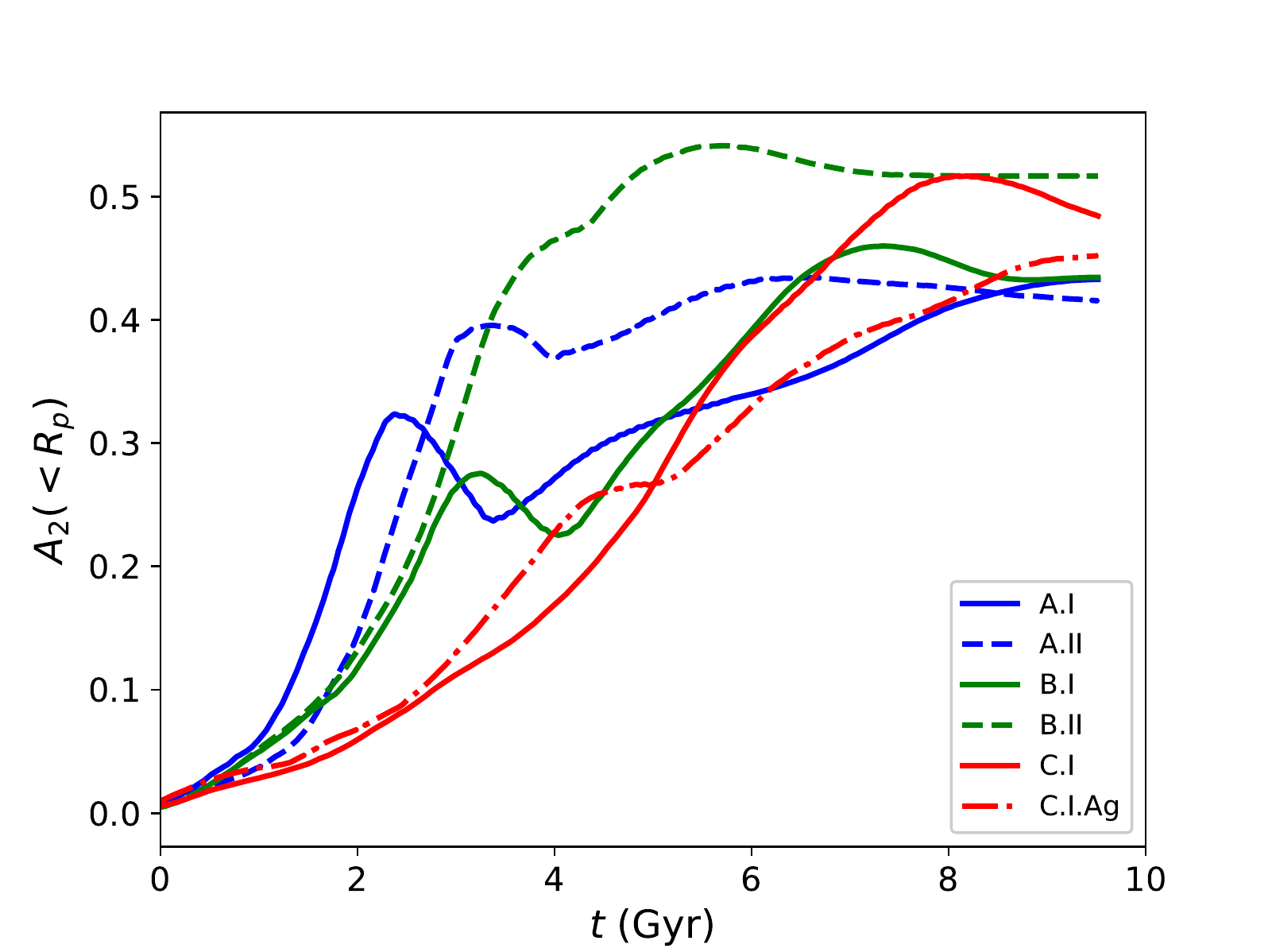}
	\caption{Mean bar strength parameter inside a cylindrical
          radius $R_p$, $A_2(<R_p)$, as a function of time. Curves are
          smoothed in time with a top-hat moving window of width
          $1\,{\rm Gyr}$.  Line colors are blue, red, and green for
          models A, B, and C, respectively.  Results for the
          fiducial runs A.I, B.I, and C.I are shown as solid curves
          while the results for the runs with high softening length,
          A.II and B.II, are shown as dashed curves.  The
          \textsc{AGAMA} model C.I.Ag is shown as a dot-dashed
          curve.} \label{fig:isolated_a2_vs_t}
\end{figure}

\begin{figure*}
	\centering
	\includegraphics[width=1.1\textwidth]{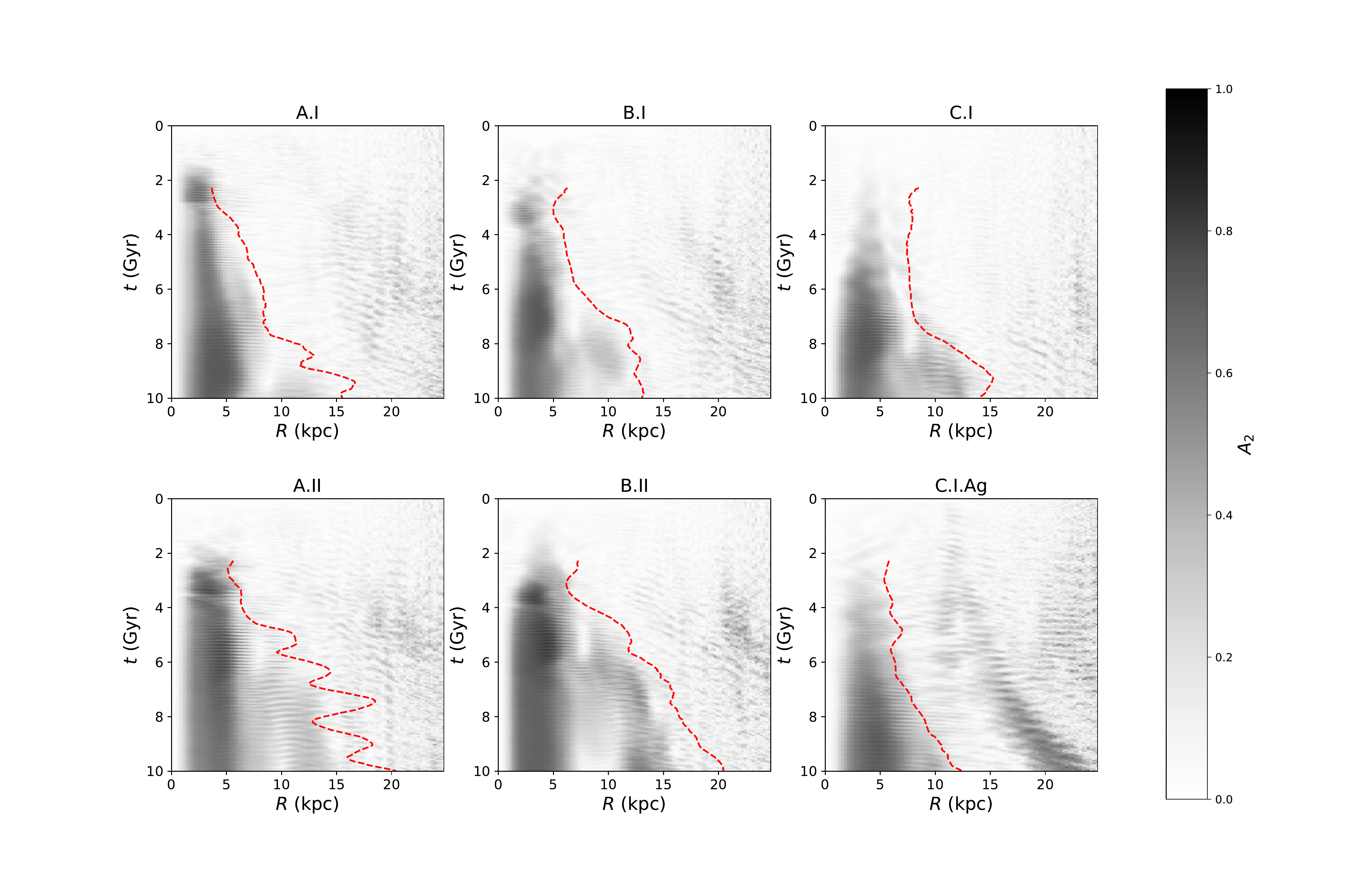}
	\caption{Bar strength parameter $A_2$ as a function of radius
          and time.  The trajectory of corotation is shown by the
          dashed red line.} \label{fig:isolated_r_t_a2}
\end{figure*}

\begin{figure*}
	\includegraphics[width=\textwidth]{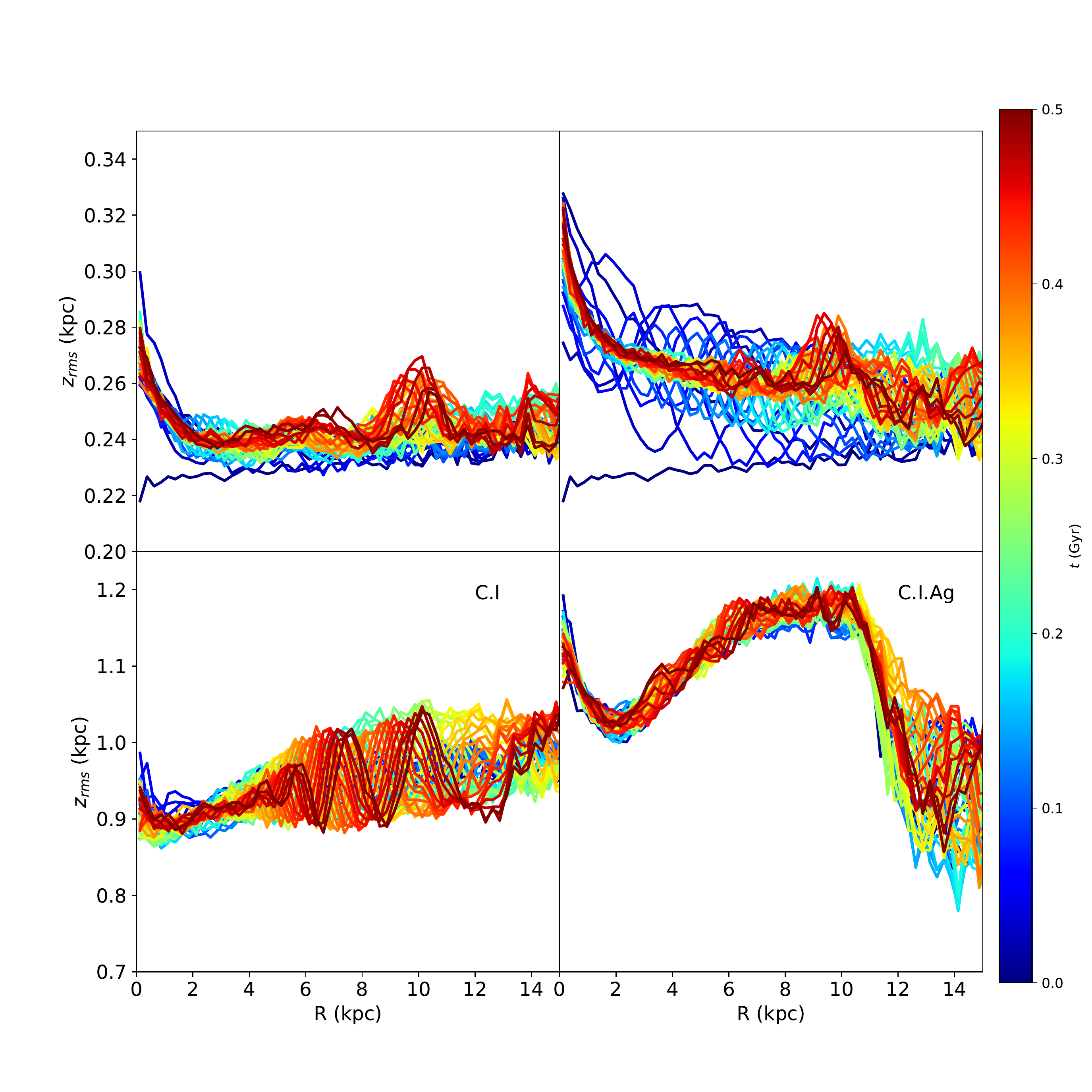}
	\caption{Root mean square height $\zrms$ as a function of
          cylindrical radius $R$ for ten snapshots equally spaced over
          the first $500\,{\rm Myr}$.  Panels are for simulations A.I
          (upper left), A.II (upper right), C.I (lower left) and
          C.I.Ag (lower right).} \label{fig:zrms}
\end{figure*}

\begin{figure*}
	\includegraphics[width=\textwidth]{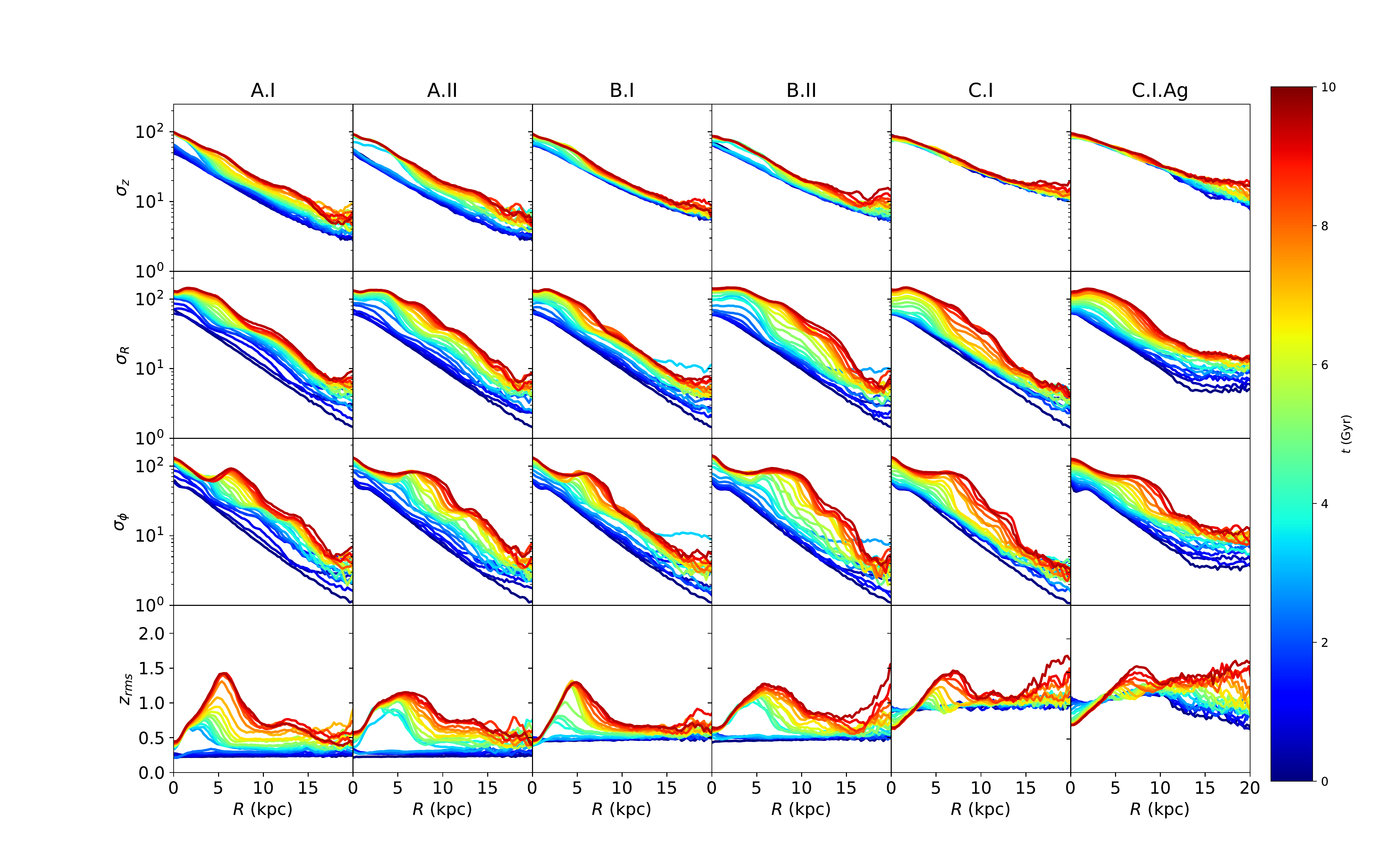}
	\caption{Diagonal components of the velocity dispersion tensor
          and $\zrms$ as a function of $R$ for different snapshots
          between $0$ and $10\,{\rm Gyr}$.  Shown, from to to bottom,
          are profiles for $\zrms$, $\sigma_z$, $\sigma_R$, and $\sigma_\phi$
          for the same size models included in
          Fig.\,\ref{fig:face_on_isolated}.} \label{fig:isolated_dispersions}
\end{figure*}

\subsection{$Q$ and $X$} 
The stability of a stellar disc is generally thought to be determined
by the Toomre-$Q$ parameter \citep{ToomreParameter}

\begin{equation} \label{eq:q}
Q \equiv \frac{\sigma_R\kappa}{3.36G\Sigma}
\end{equation}

\noindent and the Goldreich-Tremaine (swing amplification) parameter
\citep{GoldreichTremaine1978, GoldreichTremaine1979}

\begin{equation} \label{eq:xm}
X_m \equiv \frac{\kappa^2 R}{2\pi m G\Sigma}
\end{equation}

\noindent where $R$ is the Galactocentric radius of a cylindrical
$\left (R,\,\phi,\,z\right )$ coordinate system, $\Sigma$ is the
surface density of the disc, $\sigma_R$ is the radial velocity
dispersion of the disc, and $m$ is the azimuthal mode number.  The
epicyclic radial frequency $\kappa$ is given by

\begin{equation} \label{eq:kappa2}
\kappa^2 = \frac{2V_c^2}{R^2}\left (1 + \frac{d\ln{V_c}}{d\ln R}\right )
\end{equation}

\noindent where $V_c$ is the circular speed.  We assume an exponential
disc with mass $M_d$, radial scale length $R_d$, and surface density
\begin{equation} \label{eq:sigma}
\Sigma(R) = \frac{M_d}{2\pi R_d^2} e^{-R/R_d}~.
\end{equation}
\noindent Note that $\kappa$, $\sigma_R$, $\Sigma$, $V_c$, $Q$, and
$X_m$ are functions of $R$.  In what follows, we consider the radius
$R_p$ at which the contribution to the rotation curve from the disc,
$V_d$ reaches a peak value.  For an exponential disc, $R_p \simeq
2.2R_d$ and $V_{d}(R_p) \simeq 0.62 \left (GM_d/R_d\right )^{1/2}$
\citep{BT}.

Roughly speaking, $Q$ describes the susceptibility of a disc to local
instabilities.  Cold discs with low velocity dispersion and $Q<1$ are
unstable to local perturbations.  On the other hand, $X_m$ describes
the vigour with which a global perturbation with an $m$-fold azimuthal
symmetry undergoes swing amplification.  Since we are interested in
bar formation, we set $m=2$ and note that $X_2^{-1}$ is a measure of
disc self-gravity.  To see this, we use Eq.\,(4) and the expression
for $V_{d,p}$ to find

\begin{equation} \label{eq:x2}
X_2(R_p) \simeq  0.79\,\frac{V_c^2}{V_d^2}\Biggr\rvert_{R_p}
\end{equation}

\noindent where we have assumed that the logarithmic derivative in
Eq.\eqref{eq:kappa2} is zero.

For simplicity, we define

\begin{equation} \label{eq:x}
X \equiv  \frac{V_c^2}{V_d^2}\Biggr\rvert_{R_p}~.
\end{equation}

\noindent Therefore $X=2$ when the contribution of the disc to the
circular speed curve at its peak is equal to the combined
contributions of the dynamically hot components, namely the bulge and
halo.  Following \citet{EfstathiouShotNoise},
\citet{YurinSpringelStellarDisks}, use $Q_{\rm bar} = V_{\rm
  max}/\left (GM_d/R_d\right )^{1/2}$ where $V_{\rm max}$ is the
maximum circular speed.  If we assume that $V_{\rm max}\simeq
V_c(R_p)$, then $Q_{\rm bar}^2 \simeq 0.387 X$ and the stability
criterion from \citet{EfstathiouShotNoise}, $Q_{\rm bar}>1.1$, becomes
$X > 3.13$.

\subsection{Vertical Structure of Stellar Discs}

As discussed in \citet{Klypin2009} the vertical structure of a stellar
disc plays a key role in determining the properties of any bar that
forms.  In general, the vertical structure is characterized by the
vertical velocity dispersion $\sigma_z$, surface density $\Sigma$, and
scale height.  For a self-gravitating plane-symmetric isothermal disc
these quantities are connected through the relation $\sigma_z^2 =
\sqrt{12}G\Sigma \zrms$ where $\zrms$ is the root mean square distance
of ``stars'' from the midplane.  \citep{spitzer1942, camm1950}.

We can incorportate the effects of dark matter by modifying
the Poisson equation

\begin{equation} \label{eq:spitzerpoisson}
\begin{aligned}
\frac{d^2\Phi}{dz^2} & = 4\pi G \left (\rho_d(z) + \rho_h(z) \right )\\
& = 4\pi G\rho_0 \left(e^{-\Phi/\sigma_z^2} + \rho_h/\rho_0\right )
\end{aligned}
\end{equation}

\noindent where $\rho_d$ and $\rho_h$ are the densities of the disk
and halo, respectively, and $\rho_0$ is the density of the disc in the
midplane.  In the second line we assume, as is done in the pure
self-gravitating case, that the disc stars are vertically isothermal
with velocity dispersion $\sigma_z$.  We also assume that the halo
density is constant in the region of the disc.  We then solve
Eq.\,\ref{eq:spitzerpoisson} numerically.  The result is
well-described by the relation

\begin{equation}\label{eq:sigz2}
\sigma_z^2 = \sqrt{12} 
G\Sigma \zrms \left (1 + \alpha\rho_h/\rho_0\right )
\end{equation}

\noindent where the factor $1 + \alpha = 1 + \sqrt{2\pi/3}$ provides a
simple interpolation between the pure self-gravitating case and the
case where disc particles are test particles in the (harmonic)
potential of a constant density halo.  As discussed in the next
section Eq.\,\ref{eq:sigz2} holds at the 10 per cent level for our
equilibrium models.  Departures from Eq.\,\ref{eq:sigz2} might come
from radial gradients and the rotation of the disc. (See, for example,
\citet{read2014}).

Combining Eqs.\,\ref{eq:q}, \ref{eq:x}, \ref{eq:sigz2} we find 
following relation:

\begin{equation}\label{eq:r}
\frac{Q^2}{X} = 3.103 \,
\frac{\sigma_R^2}{\sigma_z^2} \,
\frac{\zrms}{R_d}\,
f\left (1 + \frac{d\ln V_c}{d\ln R}\right )~.
\end{equation}
\noindent This expression can be interpreted in several ways.  First,
if the ratios of $\sigma_R$ to $\sigma_z$ and $\zrms$ to $R_d$ are
fixed, then there is a linear relation between $Q^2$ and $X$.  On the
other hand, if one considers a family of models in which the only
variation is in the vertical structure of the disc, then the scale
height varies roughly linearly with the vertical velocity dispersion,
apart from corrections due to the contribution of the halo to the
vertical force.

\subsection{Effect of Gravitational Softening} 

Numerical effects can significantly alter the development of bars in
simulated galaxies. For example, in simulations of an isolated galaxy
that is initially in equilibrium, the onset of bar formation is
delayed when mass resolution is increased \citep{dbs2009}
essentially because the bar instability is seeded by shot noise.  The
importance of mass resolution as well as force resolution and time
stepping are also discussed in \citet{Klypin2009}.

In this section, we focus on the effects of force softening.
Equilibrium models, such as the ones used as initial conditions in
isolated galaxy simulations, satisfy the collisionless Boltzmann and
Poisson equations.  When evolved with force softening, they will begin
slightly out of equilibrium.  This effect should be most noticeable
when the softening length is comparable to or larger than the
thickness of the disc.  To gain some intuition as to this extent of
this effect we the Poisson equation in one dimension.  The potential
for a mass distribution with vertical density profile of $\rho(z)$ can
be calculated by convolving $\rho$ with the Green's function:
\begin{equation}\label{eq:greensfunction}
\Phi(z)= 4\pi G\int_{-\infty}^{\infty} \mathcal{G}(z^\prime - z)
\rho(z^\prime) \text{d} z^\prime~.
\end{equation}
For Newtonian gravity, $\mathcal{G} = |z|/2$.  For softened gravity,
we replace $\mathcal{G}$ with $\mathcal{G}_s = \frac{1}{2}\left (z^2 +
\epsilon^2\right )^{1/2}$ where $\epsilon$ is the softening length.
(The motivation for this expression is as follows: Begin with a system
of Plummer-softened particles, that is, a system where point-like
particles are replaced by particles whose spherical density profile is
proportional to $\left (r^2 + \epsilon^2\right )^{-5/2}$.  If the
particles are confined to a plane, then the vertical density profile
will be $\rho(z)\propto \left (z^2 + \epsilon^2\right )^{-3/2}$.  The
one-dimensional potential with this $\rho(z)$ is indeed proportional
to $\left (z^2 + \epsilon^2\right )^{1/2}$.)  The integral
Eq.\,\ref{eq:greensfunction} and the related integral for the vertical
force, $f(z)$, can be evaluated numerically.  As expected, the
potential energy per unit area of the system, $W \equiv \int
dz\,\rho(z) z f(z)$ is smaller than that of the same system found
assuming Newtonian gravity.  Hence, a system that is set up to be in
equilibrium under the assumption of Newtonian gravity, will be too
``warm'' for a softened gravity simulation and will ``puff up''.  To
an excellent approximation, we find that the virial ratio between the
kinetic energy per unit area and $W$ is given by $2T/W \simeq (1 +
(a\epsilon/\zrms)^2)^{b}$ where $a = 1.25$ and $b = 0.25$.  Roughly
speaking, simulations run with a softening length equal to $\zrms$
will have a virial ratio of $1.25$.

Softening may have other effects on the development of the bar.  In
principle, softening should suppress the Toomre instability on small
scales.  However, this instability develops on scales comparable to or
larger than the Jeans length, which is typically much larger than the
thickness of the disc and hence larger than the softening length for
most simulations.  On the other hand, softening may suppress buckling,
a bending instability, which is strongest on small scales.  As
discussed below, buckling appears to be responsible for regulating the
growth of bars.

\section{MODELS AND SIMULATIONS} \label{sec:thick_discs_suppress}

\subsection{Initial Conditions for Isolated Galaxy Simulations}

We follow the evolution of isolated disc-halo systems using the N-body
code \textsc{gadget-3} \citep{GadgetCodePaper}.  The initial
conditions for most of our isolated galaxy simulations are generated
with \textsc{GalactICS} \citep{GalactICS1995,WPDGalactICSReference},
which allows users to build multicomponent, axisymmetric equilibrium
systems with prescribed structural and kinematic properties.  Disc
particles are sampled from a distribution function (DF) that is a
semi-analytic function of the total energy $E$, the angular momentum
about the disc symmetry axis $L_z$, and the vertical energy $E_z = \Phi(R,0) - \Phi(R,z) + \frac{1}{2} v_z^2$, where $\Phi$ is the gravitational potential and $v_z$ is an orbit's vertical velocity.  By
design, the disc DF yields a density law in cylindrical $\left
(R,\,\phi,\,z\right )$ coordinates given, to a good approximation, by
$\rho(R,z) = \Sigma(R)\,{\rm sech}^2(z/z_d)$.  Here $\Sigma(R)$ is
exponential surface density profile (Eq.\,\ref{eq:sigma}) and $z_d$ is
the scale height. Note that $\zrms = \pi/\sqrt{12} z_d$ while the
``half-mass'' scale height used in \citet{YurinSpringelStellarDisks}
is given by $z_{1/2} \simeq 0.549z_d \simeq 0.605 \zrms$.  The disc DF
is also constructed to yield a radial velocity dispersion profile that
is exponential in $R$ with scale length $2R_d$. The halo DF is
designed to yield a truncated NFW profile \citep{NFW} as described in
\citet{WPDGalactICSReference}.

While $E_z$ used in the \textsc{GalactICS} disc DF
is conserved to a good approximation for nearly circular orbits it
varies considerable for stars that make large excursions in $R$ and
$z$.  Thus, the initial conditions for ``thick'' or ``warm'' discs
will not represent true equilibrium solutions to the dynamical
equations.  To test whether non-conservation of vertical energy
affects our results, we compare a thick disc model with
\textsc{GalactICS} initial conditions with a similar one where the
initial conditions are generated with \textsc{AGAMA} \citep{agama}.
In principle, this action-based code should yield initial conditions
that are closer to a true equilibrium system than ones based on $E_z$
especially for thick discs.

\subsection{Description of simulations}

In this section, we describe a suite of simulations where $Q$ and $X$
are fixed and where the velocity dispersion and scale length
ratios are allowed to vary.  Our aim is to test the hypothesis that scale
height plays a key role in the development of bars.  The parameters
for our simulations are summarized in Table \ref{tab:sims}.  Our suite
of isolated galaxy simulations form a sequence A, B, C in increasing
thickness.  The models have the same rotation curve decomposition,
which is shown in the top panel of Fig. \ref{fig:rcs}.  By design, the
contribution to the rotation curve from the disc is slightly below
that of the halo at $R_p$.  Therefore our models have $X$ slightly
greater than $2$ and should be susceptible to global instabilities.

The fiducial simulations are run with a softening length of
$184,\,{\rm pc}$, which is about two thirds of the scale height of our
thinnest model (A.I).  The simulations A.II and B.II use a softening
length of $736\,{\rm pc}$, which is close to the value assumed in
\citet{YurinSpringelStellarDisks}.  The simulation C.I.Ag is similar
to C.I (large scale height) but run with \textsc{AGAMA} initial
conditions.  A comparison of its rotation curve decomposition with
that for model C.I is shown in the top panel of Fig. \ref{fig:rcs}.
The contributions from the disks in the two models are nearly the same
and the contributions from the halos differ significantly only beyond
$\sim 10\,{\rm kpc}$.  The simulations A.III and B.III use a scheme to
isotropize vertical motions and effectively shut off buckling and are
discussed in \S \ref{sec:buckling}.

In addition to these isolated galaxy simulations we run two
cosmological simulations using the disc insertion scheme described in
\citet{Bauer2018a}.  The initial conditions for these models, labeled
D.I and E.II, are identical except for the vertical scale height and
softening length, which are larger in E.II.  Thus, these models are
cosmological analogs to A.I and B.II.  The rotation curves for these
models are shown in the bottom panel of Fig. \ref{fig:rcs}.  The
models themselves are discussed in Section \S \ref{sec:cosmo}.

\subsection{Comparison with Previous Work}

While the parameters $Q$ and $X$ allow one to predict the rapidity and
vigour with which instabilities develop in disc galaxies that are
actually imperfect predictors of the strength and length of bars at
late times.  The point is illustrated in \citet{WPDGalactICSReference}
where results for a suite of 25 simulations that explore the $Q-X$
plane are presented.  By design, the initial conditions for the models
satisfy observational constraints for the Milky Way such as the
rotation curve, the local vertical force, and the velocity dispersion
toward the bulge.  (See \citet{hartmann2014} for a further analysis of
these simulations.)  As expected, the onset of the bar instability is
delayed in models with large initial values for $Q$ and/or $X$.
However, the dependence on these parameters of the bar strength and
length is more complicated.  In Fig.\ref{fig:qxa2} we show the bar
strength parameter $A_2$ and length of the bar across these models.
Evidently, the models that form the strongest and longest bars have
intermediate values of $Q$ and $X$.  The implication is that models
where the instabilities grow too quickly lead to weaker and somewhat
shorter bars.  Bar formation appears to be a self-regulating process.

Table \ref{tab:sims} gives the relevant parameters for the eight
disc-halo models from \citet{YurinSpringelStellarDisks} as well as the
disc-bulge-halo model for M31 from \citet{gauthier2006}.  In the
\citet{YurinSpringelStellarDisks} simulations discs are inserted into
dark matter haloes from the cosmological Aquarius simulation.  In this
respect, they are similar to the disc-insertion simulations described
in Section \S \ref{sec:cosmo}.  The initial discs in these models all
have a scale height to scale length ratio of $0.2$ and a radial to
vertical velocity dispersion ratio of $1$.  As discussed above, these
choices mean that their discs were chosen from a one-parameter family
of models within the $Q-X$ parameter space.

\section{ISOLATED GALAXY SIMULATIONS}\label{sec:isolated}

\subsection{Morphology of Bar Forming Galaxies}

Face on surface density maps for models A.I, A.II, B.I, B.II, C.I, and
C.I.Ag are shown in Fig.\,\ref{fig:face_on_isolated}.  All discs form
bars by the end of the simulation ($t=10\,{\rm Gyr}$).  However, bar
formation appears to be delayed in models B.I and C.I relative to that
in model A.I while the final bar in A.I is shorter than those in B.I
and C.I.  Other $m=2$ features are also evident.  These include
two-armed spiral structure, most clearly seen in A.I and B.III and
elliptical rings, as, for example, in C.I.

Evidently, the dominant mode for in-plane perturbations is $m=2$
Nevertheless, there are strong $m=3$ structures in the $1.5\,{\rm
  Gyr}$ snapshot of the A.I and A.II simulations and hints of a weak
$m=3$ structure in the same snapshot of B.II.

A larger softening length seems to lead to stronger bars at
intermediate times.  We see this in the comparison of A.I and A.II or
B.I and B.II in the $3.0\,{\rm Gyr}$ and $4.5\,{\rm Gyr}$ snapshots.

\subsection{Bar strength parameter $A_2$}

It is convenient to think of the azimuthal distribution of particles
in a given radial bin as a Fourier series.  We define the coefficient
of the Fourier component with $m$-fold azimuthal symmetry to be
\begin{equation}\label{eq:cm}
c_m =  \frac{1}{M_S} \sum_{j \in S} \mu_i e^{i m \phi}
\end{equation}

\noindent where $\mu_i$ is the mass of the $i$-th particle and $S$ is
a circularly symmetric region of the disc.  The normalization is
chosen so that a distribution of particles along a line through the
origin will have $|c_m|=1$ for all $m$ even.  Moreover, for a uniform
distribution of particles, $c_0=1$ and $c_m=0$ for all $m>0$.  The
amplitude and phase for the $m$-th Fourier coefficient are given by
$A_m \equiv \vert c_m \vert$ and $\phi_m = \text{arg } c_m$,
respectively.  Note that both of these quantities depend on the region
$S$

Fig. \ref{fig:isolated_a2_vs_t} shows a plot of the mean $A_2$ inside
the radius $R_p$ as a function of time for the fiducial simulations,
the two simulations with high softening, and the thick disc simulation
with initial conditions from \textsc{agama}.  Consider first the
fiducial (low-softening) simulations.  Initially, $A_2$ grows roughly
exponentially with a growth rate that decreases with increasing
thickness.  In simulations A.I and B.I, the end of exponential growth
is followed by a decrease in $A_2$ after which $A_2$ again increases,
now, approximately linearly with time.  In the thick disc case (C.I)
exponential growth transitions directly to linear growth.  The trend
is for exponential growth to end at later times as one goes to thicker
discs.  It is worth noting that the value of $A_2$ at $10\,{\rm Gyr}$
is similar in the three low-softening simulations.

In the thin disc case, an increase in softening appears to delay the
onset of exponential growth as well as the time at which exponential
growth ends.  Furthermore, the drop in $A_2$ is less severe.  Though
the value of $A_2$ at the end of the simulation is approximately the
same in the low and high softening cases, the bar strength, as
measured by $A_2$ is larger in the high-softening case at intermediate
times between $4$ and $8\,{\rm Gyr}$.  For the intermediate thickness
case (B.I and B.II) softening has little effect on the initial
growth rate of $A_2$.  But as in the thin disc case, softening
allows exponential growth to continue to later times and the final bar
is about twenty per cent stronger as compared with the low-softening
case.  Once again we see that the effect of high softening is to
produce stronger bars at intermediate times.

The evolution of $A_2$ for the thick disc runs with \textsc{GalactICS}
and \textsc{agama} initial conditions are fairly similar.  In particular,
the initial growth rate is almost identical as are the final values.

Fig. \ref{fig:isolated_a2_vs_t} encapsulates bar strength into a
single number, the mean $m=2$ Fourier amplitude inside $2.2$ disc
scale lengths, or about $5.5\,{\rm kpc}$.  A more complete picture of
bar strength is presented in Fig.\,\ref{fig:isolated_r_t_a2} where we
plot $A_2$ as a function of $R$ and $t$. The figure is constructed by
calculating $c_2$ (Eq.\,\ref{eq:cm}) for cylindrical rings of radius
$200\,{\rm pc}$.  Also shown is the corotation radius, which is
determined from the pattern speed $\Omega_p$ and rotation curve. The
former is given by a numerical estimate of ${\text{d}
  \phi_2}/{\text{d} t}$; corotation is found by determining the radius
at which $\Omega_p = V_c/R$.  Thus, since our galaxy models have
roughly flat rotation curves beyond $5\,{\rm kpc}$, the corotation
essentially gives the inverse patter speed or pattern period.

From Fig. \ref{fig:isolated_a2_vs_t} we see that the corotation radius
tends to grow with time and provides an envelope for the bar and other
$m=2$ structures such as two-armed spirals and elliptical rings.  The
bar pattern speed is therefore decreasing with time, presumably due to
dynamical friction between the bar and both the disc and dark halo
\citep{debattista1998, debattista2000}.  It is worth noting that the
corotation radius increases more rapidly in simulations with high
softening.  The naive interpretation is that softening somehow
increases the frictional coupling between the bar and disc or halo
particles.  A more likely explanation is that with a high softening
length comes stronger bars.  Since the acceleration on the bar due to
dynamical friction scales as the mass of the bar, stronger bars should
spin down more rapidly.

As in Fig. \ref{fig:isolated_a2_vs_t} we see that bar formation
is delayed in models with thicker discs.  Bar formation is 
well underway by $2\,{\rm Gyr}$ in A.I but doesn't really take hold
until $4\,{\rm Gyr}$ in C.I.  Moreover, the first hints of $m=2$ power
in C.I arise further out at radii closer to $5\,{\rm kpc}$.  

The dip in bar strength is clearly seen between $2.5-3\,{\rm Gyr}$ in
A.I and between $3.5-4\,{\rm Gyr}$ in B.I.  As discussed below, we
attribute this dip to buckling.

\subsection{Vertical Structure and Velocity Dispersion}

Figs.\,\ref{fig:zrms} and \ref{fig:isolated_dispersions} show the
$\zrms$ and velocity dispersion profiles for a sequence of snapshots
in various models.  The first of these plots focuses on the evolution
of $\zrms$ during the initial $500\,{\rm Myr}$ of the simulation.  The
top panels show the $\zrms$ profiles for simulations A.I and A.II and
illustrate the effect softening has on the evolution from
``equilibrium'' initial conditions.  As discussed in Section 2 as
system that is initialized to be in equilibrium under the assumption
of Newtonian gravity will be out of equilibrium if evolved with
softened gravity.  In particular, the mean potential energy will be
systematically low and the system will puff up.  For our thin disc
model, $\zrms\simeq 230\,{\rm pc}$.  In the high softening case,
$\epsilon = 736\,{\rm pc} \simeq 2.2\zrms$, we estimate the virial ratio
for the vertical structure to be $2T/W\simeq 1.7$.  Of course, the
excess kinetic energy will redistribute itself into both kinetic and
potential energy.  The upshot is that the system quickly settles into
a new state with a thickness somewhat larger than the initial one as
seen in the right hand panel.

The bottom panels in Fig.\,\ref{fig:zrms} provide a comparison of
$\zrms$ profiles for the thick disc simulations with
\textsc{GalactICS} and \textsc{agama} initial conditions.  We first
note that $\zrms$ is approximately constant in the C.I but varies by
about $200\,{\rm pc}$ in C.I.Ag.  This difference is simply a
reflection in how the initial conditions are set up.  In both cases,
the scale height depends implicitly on the functional form of the DFs,
which are written in terms of either $E,\,E_z,\,$ and $L_z$ or the
actions.  The \textsc{GalactICS} case does exhibit transient wavelike
perturbations with a peak to trough amplitude of $100\,{\rm pc}$ at
radii $R>4\,{\rm kpc}$.  A plausible explanation for these
oscillations is that they are due to the fact that $E_z$ is not a true
constant of motion.  In any case, the system quickly settles to to a
new equilibrium state not too different from the initial one.

Fig.\,\ref{fig:isolated_dispersions} shows profiles for $\zrms$ and the
diagonal components of the velocity dispersion tensor, $\sigma_z$,
$\sigma_R$, and $\sigma_\phi$.  The effect of bar formation is readily
apparent in the $\zrms$ and $\sigma_z$ profiles.  In simulation A.I,
for example, bar formation, which begins around $t\simeq 1.5\,{\rm
  Gyr}$ is accompanied by thickening and vertical heating.  By the end
of the simulation, $\zrms$ increases linearly with $R$ from a
central value of about $400\,{\rm pc}$ to $1.5\,{\rm kpc}$ at a radius
of about $6-7\,{\rm kpc}$ and then decreases beyond this radius.  The
evolution is similar in simulations B.I and C.I.  Interestingly
enough, the central and peak values are very similar in all three
cases even though the initial thick of the discs are very different.
Indeed, the central value for $\zrms$ actually decreases with time
in our thick disc simulation.  

Vertical heating of the disc in the central regions also seems to be
connected with bar formation, at least in the thin and intermediate
thickness cases.  In A.I, for example, the central velocity dispersion
appears to increase rapidly starting around $3\,{\rm Gyr}$ and
reaching a final value of $\sim 100\,{\rm km\,s^{-1}}$, which is
roughly a factor of two larger than the initial value.  As with
$\zrms$, the value of the central vertical velocity dispersion is
nearly identical in all models.  Evidently, the final vertical
structure of the barred disc is insensitive to initial conditions.

All models show significant in-plane disc heating across the disc and
throughout the simulation.  While the initial radial velocity
dispersion profile is exponential in $R$ the final profile is almost
flat within the central $5\,{\rm kpc}$.  Thus, the greatest increase
in radial velocity dispersion is at about this radius, which
corresponds to the end of the bar.  On the other hand, the greatest
increase in $\sigma_\phi$ occurs at larger radii, closer to $10\,{\rm
  kpc}$.

\subsection{Simulations Where Buckling is Suppressed}\label{sec:buckling}

Buckling is a well-known phenomena often seen in simulations of
bar-forming galaxies where the bar bends in and out of the disc plane.
Eventually, these coherent oscillations are converted to random
vertical motions \citep{BT}.  Buckling typically leads to shorter and
weaker bars \citep{VP2004}

To isolate the effects of buckling we implement a simple scheme
that prevents the instability from taking hold.  Essentially, at
each timestep, we reverse the vertical components of the position,
velocity, and acceleration for a fraction $p$ of disc particles.
In practice, we choose $p=0.25$ though the results are insensitive
to the exact value.

\begin{figure}
	\includegraphics[width=0.5\textwidth]
{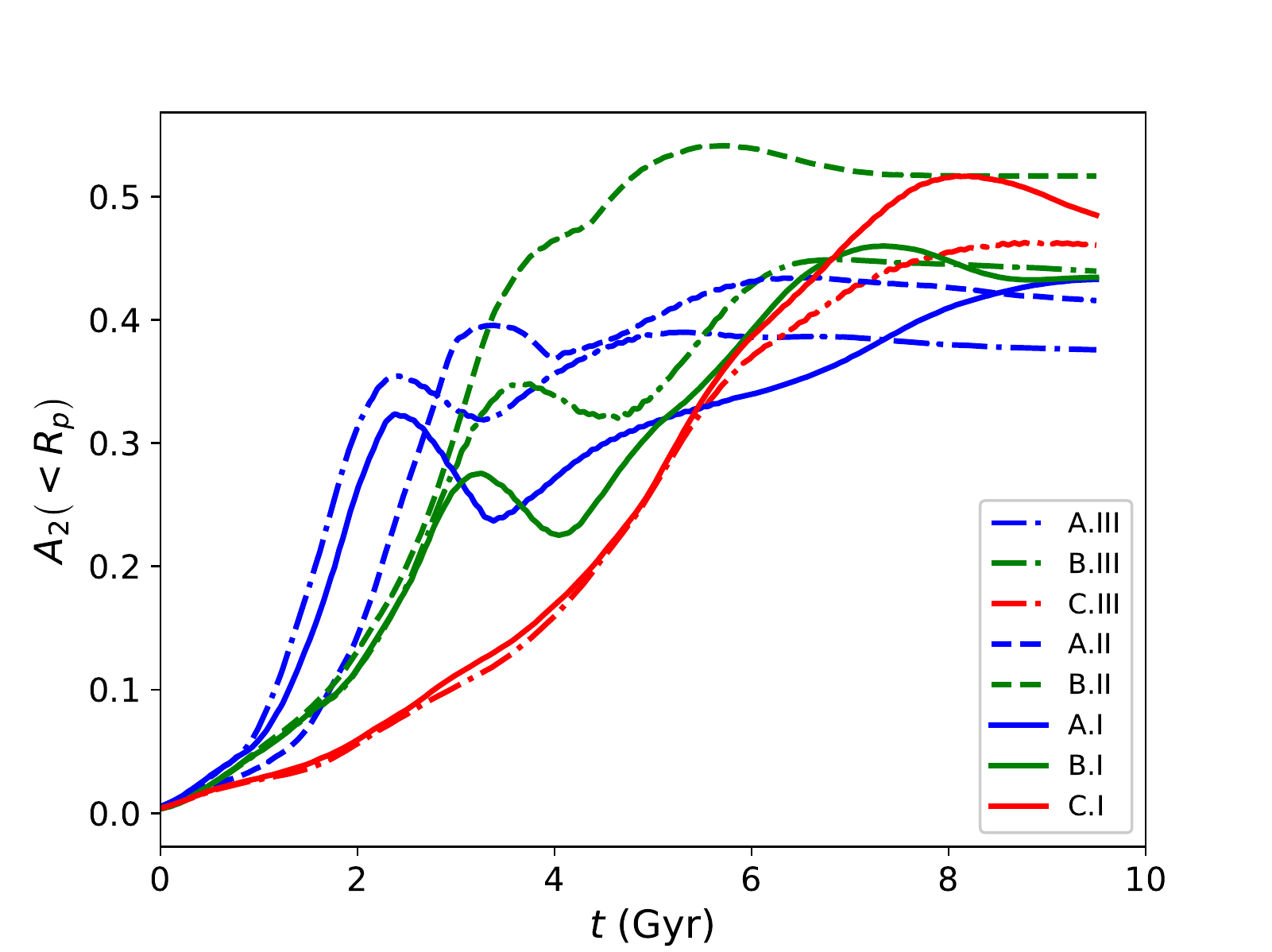}
	\caption{Mean bar strength parameter inside the cylindrical
          radius $R_p$, $A_2(<R_p)$, as a function of time.  The
          figure is essentially the same as
          Fig.\,\ref{fig:isolated_a2_vs_t} though this time we include
          simulations A.III, B.III, and C.III where buckling is
          suppressed.} \label{fig:isolated_a2_vs_t_no_buckle}
\end{figure}

Fig. \ref{fig:isolated_a2_vs_t_no_buckle} shows the effect suppressing
buckling has on the disc evolution.  In the thin disc case, the bar
instability develops a bit faster when buckling is suppressed.  More
importantly, the drop in $A_2$ seen in simulation A.I is not as strong,
thus confirming the notion that buckling regulates the strength of
bars.  Buckling has a similar effect on our intermediate thickness
runs.  Furthermore, the effect of suppressing buckling is similar, in
some respects, to the effect of increasing softening as can be seen by
noting similarities between A.II and B.III.  Finally, we note that
buckling doesn't appear to occur in our thick disc simulations.

\begin{figure*}
	\includegraphics[width=\textwidth]{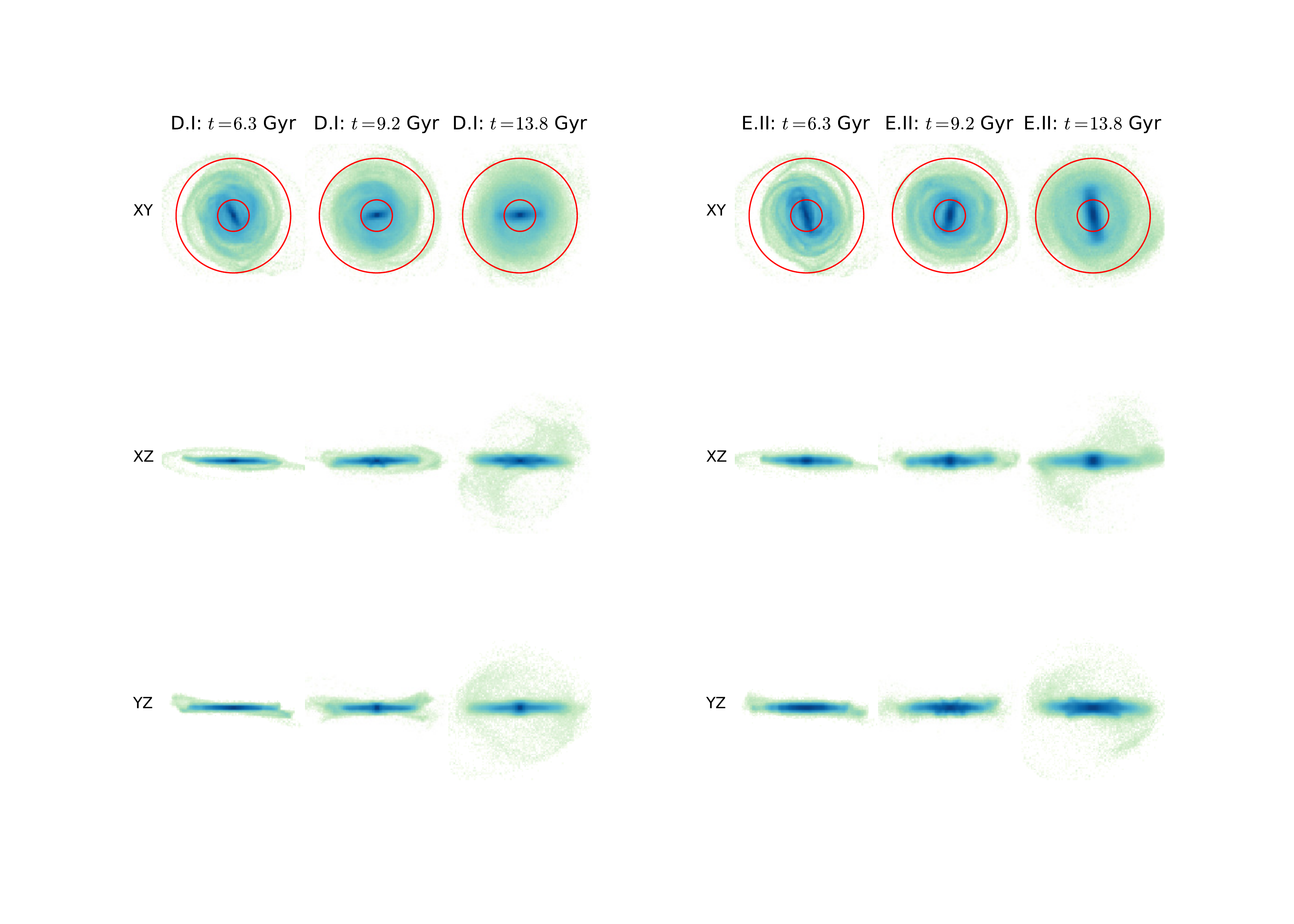}
	\caption{Projections for the D.I (left three columns) and E.II
          (right three columns).  The three columns for each
          simulation correspond,from left to right, to $2.2\,{\rm
            Gyr}$, $5.9\,{\rm Gyr}$, and $13.7\,{\rm Gyr}$ after the
          Big Bang. The overlaid red circles have radii $R_p$ and 20
          $h^{-1} \,$kpc.} \label{fig:face_on_cosmo}
\end{figure*}

\begin{figure}
	\includegraphics[width=0.5\textwidth]{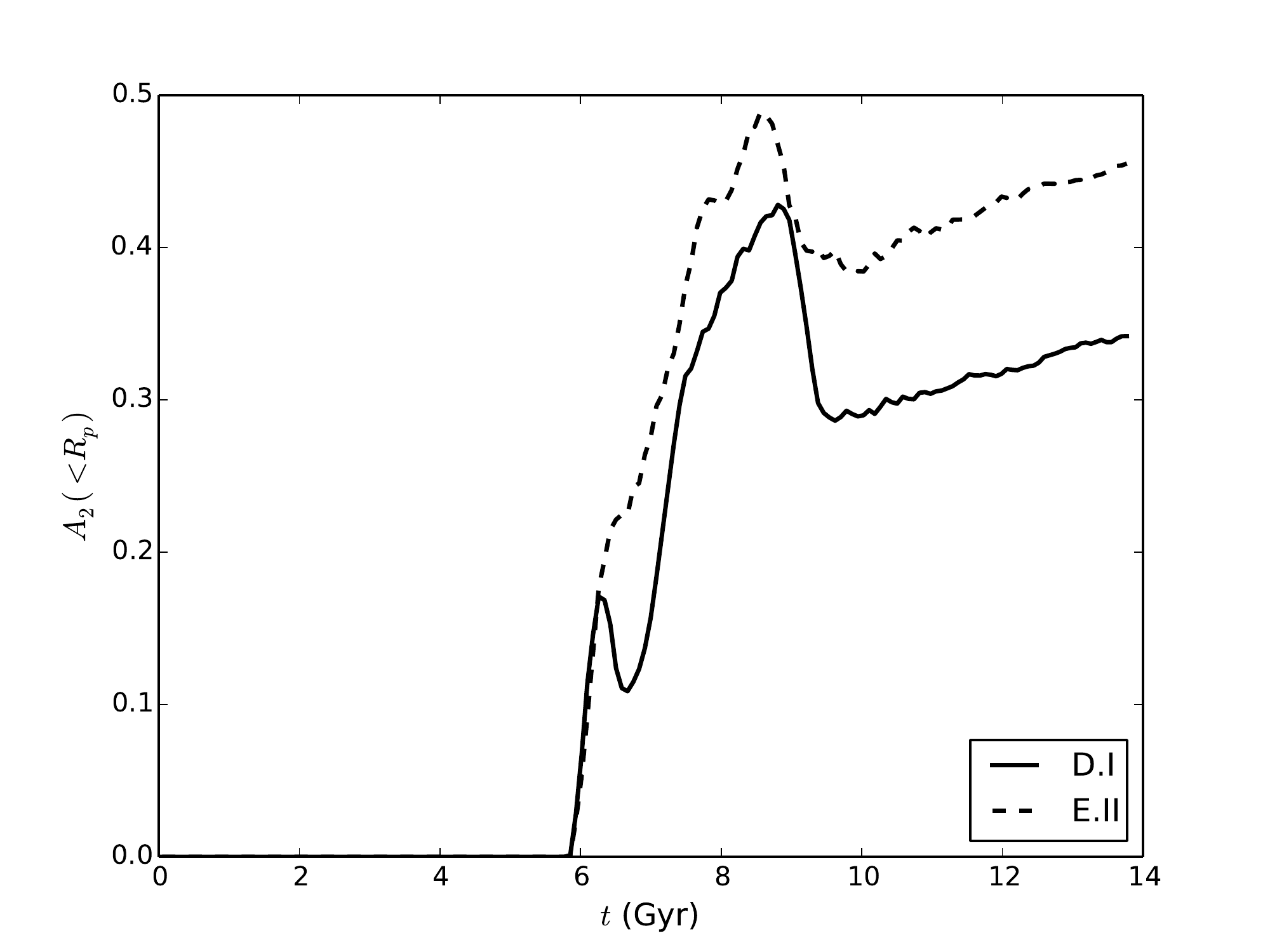}
	\caption{$A_2(<R_p)$ as a function of the age of the Universe
          for simulations D.I (solid curve) and E.II (dashed
          curve). } \label{fig:cosmo_a2_vs_t}
\end{figure}

\begin{figure}
	\includegraphics[width=0.5\textwidth]{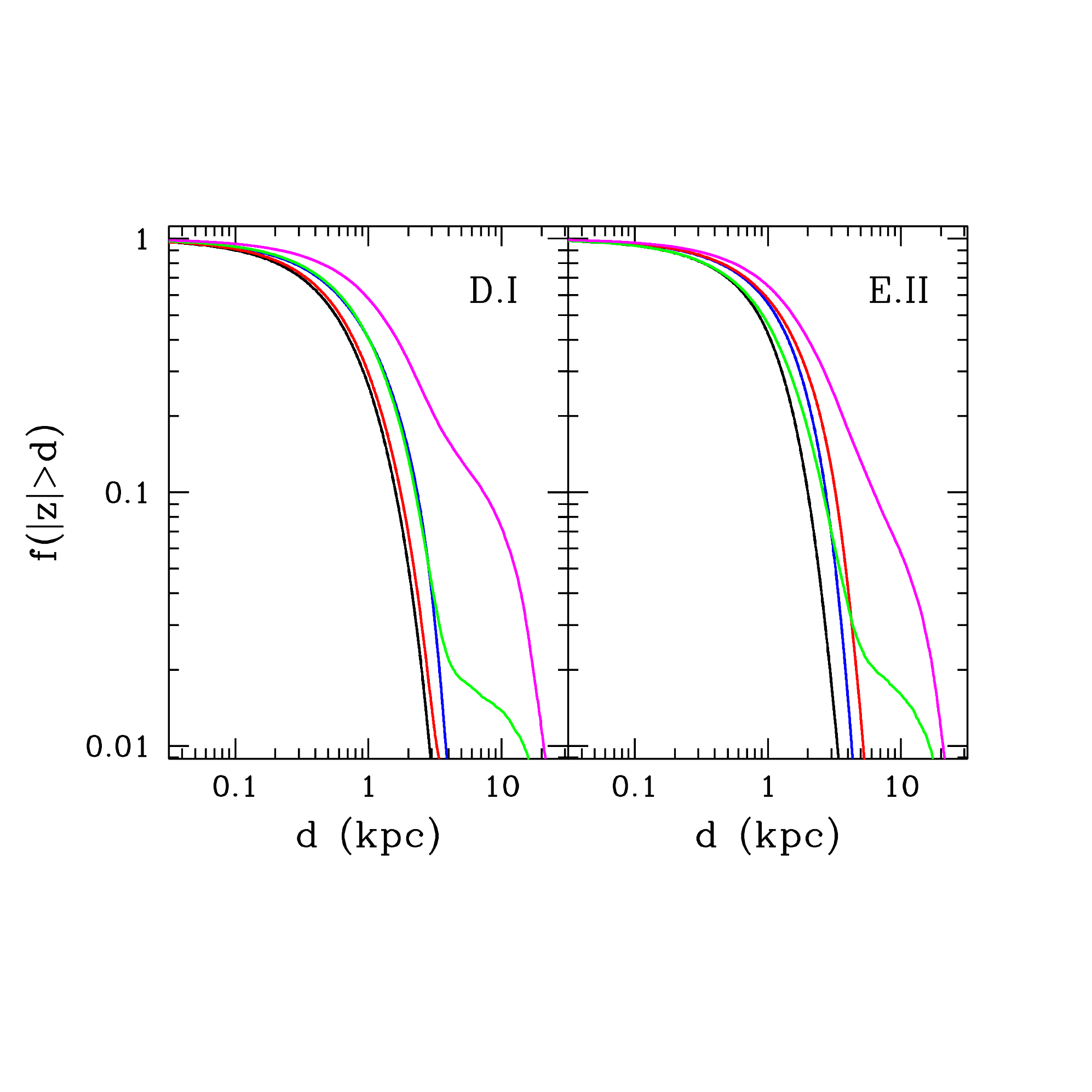}
	\caption{Fraction of particles with distance from the midplane greater
than some distance $d$ as a function of $d$.  The difference colours correspond to
different bins in cylindrical radius $R$: $0<R<5\,{\rm kpc}$ --- black;
$5\,{\rm kpc} < R < 10\,{\rm kpc}$ --- blue;
$10\,{\rm kpc} < R < 15\,{\rm kpc}$ --- red;
$15\,{\rm kpc} < R < 20\,{\rm kpc}$ --- green;
$20\,{\rm kpc} < R\,{\rm kpc}$ --- magenta.}
 \label{fig:kicked_up_disc}
\end{figure}

\section{COSMOLOGICAL SIMULATIONS} \label{sec:cosmo}

Disc galaxies simulated from axisymmetric, equilibrium initial
conditions, as was done in the previous section, form bars at rates
and with strengths that depend on their intrinsic scale height of the
disc and on the force resolution of the simulation. In this section,
we investigate the extent to which these results hold in a
cosmological environment.  In particular we follow the evolution of a
thin disc with moderate softening and a thick disc with high softening
that are embedded in identical cosmological haloes.

\subsection{Simulation Setup; Inserting Discs into Cosmological Haloes}

We model a stellar disc in a cosmological halo using the disc
insertion scheme described in \citet{Bauer2018a}.  This scheme, which
builds on the methods developed by
\citet{BerentzenShlosmanStellarDisks}, \citet{DeBuhrStellarDisks}, and
\citet{YurinSpringelStellarDisks} uses an iterative procedure to
initialize the disc.  The first step is to run a pure dark matter
simulation and identify a suitable halo.  The system is then rerun
from redshift $z_g$ to $z_l$, this time with a disc potential that
grows slowly in mass and radius.  Doing so allows the halo particles
to respond to the gravitational field of the would-be disc.  At $z_l$,
the rigid disc is replaced by an N-body system and the ``live'' disc-halo
system is evolved to the present epoch.

For our pure dark matter simulation, we implement the zoom-in
technique of \citet{KatzQuasarZoom} and \citet{NavarroWhiteZoom},
broadly following the recommendations of \cite{onorbe_etal_2014},
which allows us to achieve very high spatial and mass resolution for a
single halo while still accounting for the effects of large-scale
tidal fields.  We choose cosmological parameters based on the results
from Planck 2013 \citep{planck_2014} with $H_0=67.9\,{\rm
  km\,s}^{-1}\,{\rm kpc}^{-1}$, $\Omega_b = 0.0481$, $\Omega_0 =
0.306$, $\Omega_\Lambda = 0.694$, $\sigma_8 = 0.827$, and $n_s =
0.962$.  N-body initial conditions for the dark matter particles are
generated with the \textsc{music} code \citep{music}.  
We select a suitably-sized halo for a Milky Way-like galaxy, namely
one with a $z=0$ mass of $1.23\times 10^6\,h^{-1} M_\odot$
that comprises $10^6$.

During its growth phase from $z_g=3$ to $z_l=1$, the disc is treated
as a rigid body whose orientation and center-of-mass position evolve
according to the standard equations of rigid body dynamics.  At
$z_l$, we swap a live disc for the rigid one using the
\textsc{GalactICS} code
\citep{KGGalactICSReference,WPDGalactICSReference}, which generates a
three-integral DF disc in the best axisymmetric approximation to the
halo \citet{Bauer2018a}.

We run two simulations, D.I, which assumes a thin disc with a
softening length of $184\,{\rm pc}$ and E.II, which assumes a thick
disc with a softening length of $736\,{\rm pc}$.  The softening length
chosen for D.I is in accord with the criteria outlined in
\citet{power_et_al_2003}.  The simulations D.I and E.II roughly
correspond to A.I and B.II, respectively.  As well, E.II is similar to
the discs considered by \citet{DeBuhrStellarDisks},
\citet{YurinSpringelStellarDisks} and \citet{Bauer2018a}, whereas D.I
is more consistent with typical discs considered in isolated galaxy
suites like \citet{WPDGalactICSReference}.

\subsection{Results}

Results from our two cosmological simulations are displayed in
Figs.\,\ref{fig:face_on_cosmo} and \ref{fig:cosmo_a2_vs_t}.  The
former shows projections of the mass density at three epochs while the
latter gives $A_2(<R_p)$ as a function of time.  Evidently, the discs
in both cases roughly follow the same evolutionary sequence that was seen in
the isolated galaxy simulations: rapid growth of the bar strength
followed by a period where the bar strength decreases,
presumably due to buckling, and finally steady strengthening of the
bar.  The three epochs chosen in Fig.\,\ref{fig:face_on_cosmo}
correspond to the initial growth phase of the bar
($a=0.6,~t=6.3\,{\rm Gyr}$), an epoch after buckling
($a=0.7,~t = 9.2\,{\rm Gyr}$), and the present epoch at
$t=13.8\,{\rm Gyr}$.  Visually, the bar appears to be stronger and
longer in the E.II run than D.I one at each of these epochs but
perhaps most notably in the final one.  Indeed, the disc in E.II looks
very similar to those seen in the simulations of
\citet{DeBuhrStellarDisks}, \citet{YurinSpringelStellarDisks}, and
\citet{Bauer2018a}.  The fact that the bar in D.I is weaker than the
one in E.II is consistent with the results from our isolated galaxy
simulations that thicker discs produce stronger bars (See
Fig.\,\ref{fig:face_on_isolated}.

The most significant difference between bar formation in the
cosmological setting and bar formation in isolated galaxies concerns
the initial growth of the bar.  For isolated galaxies,
Fig.\,\ref{fig:isolated_a2_vs_t} clearly shows that the onset of bar
formation is delayed for thicker discs.  Conversely, in the
cosmological case, $A_2$ rapidly grows to a value of $\sim 0.17$
within the first few hundred Myr after the disc ``goes live''
regardless of the disc thickness.  At this point, the bar in the thin
disc model decreases in strength with $A_2$ dropping to $\sim 0.11$
before resuming its growth.  By contrast, the bar in the thick disc
model continues to grow monotonically.  As in the isolated galaxy
simulations, self-regulating processes such as buckling are more
efficient in the thin disc case and so $A_2$ in simulation D.I lags
behind that of E.II.  We note that in both cases, $A_2$ drops
significantly at around $t=7.5\,{\rm Gyr}$ and grows steadily
thereafter.

Our interpretation of these results is as follows: In isolation, where
discs start from axisymmetric initial conditions, the only source of
the $m=2$ perturbations that drive bar formation is shot noise from
the N-body distribution.  Evidently, making a disc thicker slows the
growth of these perturbations.  On the other hand, $m=2$ perturbations
abound in the cosmological environment where halos are clumpy and
triaxial.  The initial growth of the bar may, in fact, be relatively
insensitive to the thickness of the disc, once discs are placed in a
cosmological setting.  On the other hand, disc thickness does effect
the resilience of the bar to self-regulating processes, such that
buckling and therefore thick discs tend to have stronger bars.

Finally, we note that in both D.I and E.II, a significant number of
particles are found at high galactic latitudes.  These particles
represent stars ``kicked-up'' from the disc presumably by the
large-scale tidal fields of the halo and interactions between the disc
and halo substructure.  Kicked-up stars have been seen in cosmological
simulations by \citet{purcell2010}, \citet{mccarthy2012} and
\citet{tissera2013}.  Their existence was inferred in a combined
analysis of kinematic and photometric data for the Andromeda galaxy
\citep{dorman2013}.  Furthermore, the idea of kicked-up stars has been
invoked by \citep{pricewhelan2015} to explain the Triangulum-Andromeda
stellar clouds \citep{rochapinto2003, martin2014} and by
\citep{sheffield2018} to explain the Monoceros Ring \citep{yanny2000,
  newberg_2002} and associated A13 stellar overdensity
\citep{sharma2010}.

In Fig.\,\ref{fig:kicked_up_disc} we show the fraction of stars with
$|z|>d$ for different regions of the discs in our two cosmological
simulations.  The results are strikingly similar for the two
simulations as is already evident from a visual inspection of 
Fig.\,\ref{fig:face_on_cosmo}. The implication is that
the processes by which stellar orbits are
perturbed out of the disc plane are relatively insensitive to the
vertical structure of the disc.  We see that very few of the stars
with cylindrical radius $R<15\,{\rm kpc}$ and only $1-2\,\%$ of the
stars between $15$ and $20\,{\rm kpc}$ are kicked-up to distances
greater than $3\,{\rm kpc}$ though some stars from the $15-20\,{\rm
  kpc}$ region do end up with $|z|> 10\,{\rm kpc}$.  On the other
hand, $20\,\%$ of the stars from the region beyond $20\,{\rm kpc}$ end
up with $|z|> 3\,{\rm kpc}$ from the midplane and $10\,
10\,{\rm kpc}$.  Of course, the actual number of stars is certainly
larger since a fraction of the kicked-up stars will be passing through
the disc with large vertical velocities.

\section{CONCLUSIONS}\label{sec:conclusions}

The seminal work of \citet{PeeblesOstriker1973} introduced the notion
that disc dynamics provides a powerful constraint on the structure of
discs and the halos in which they reside.  In short, discs that are
dynamically cold and that account for a substantial fraction of the
gravitational force that keeps their stars on nearly circular orbits
are unstable to the formation of strong bars and spiral structure.
The existence of galaxies with weak bars or no bars at all tells us
that at least some discs are relatively low in mass (i.e., submaximal)
and/or dynamically warm.

The theoretical analysis presented in Section 2 showed with a few
simple assumptions (e.g., exponential surface density profile) one can
derive a relation among the structural parameters of a disc in
approximate equilibrium and thus a constraint on initial conditions
that one might choose for simulations.  For example, if one fixes
$h_d/R_d$ and $\sigma_R/\sigma_z$, as was done in
\citet{YurinSpringelStellarDisks}, then there is an approximately
one-to-one relationship between $Q$ and $X$.  Likewise, fixing $Q$ and
$X$ implies a relationship between $h_d/R_d$ and $\sigma_R/\sigma_z$.
These results have important implications for applying disc dynamics
as a constraint on models of galaxy formation.  In particular,
inconsistencies between bar demographics in a galaxy formation model
and in observational surveys may reflect differences in the scale
height and vertical velocity dispersion of model and real galaxies.

One lesson from our work and the work of others is that the relation
between structural parameters of galaxies and bar strength and length
is often rather complicated.  This observation is no doubt due, at
least in part, to the self-regulating nature of bar formation.  When
bars develop rapidly, they tend to buckle, which leads to weaker and
shorter bars \citep{VP2004}.  Thick discs appear to be more resilient
to buckling, which may explain why bars in these models often end up
stronger and longer than bars in thin-disc models \citep{Klypin2009}.
For similar reasons, gravitational softening can affect the
development and ultimate strength of bars.

In simulations of isolated galaxies from ``pristine'' equilibrium
initial conditions, bar formation is seeded by the shot noise of the
N-body distribution.  On the other hand, bars in a cosmological
environment are subjected to large perturbations including the $m=2$
ones that drive bar formation.  Thus, the fact that bar formation is
delayed in thick disc models of isolated galaxies may be purely
academic --- bar formation in the cosmological environment will be
initiated by a variety of stochastic effects regardless of the
thickness of the disc.  On the other hand, the resilience of thick
disks to buckling {\it is} relevant in the cosmological setting and
may explain why thick disks tend to form strong bars.  The upshot is 
that a proper understanding the distribution of bars in cosmological models
must go hand-in-hand with a proper understanding of the vertical 
structure of discs.

Clearly, a more exhaustive exploration of the model parameter space is
in order.  One might, for example, include galaxy scaling relations to
further constrain the space of models.  In addition, it would be of
interest to insert different discs (and for that matter, nearly
identical ones) into different halos in order to explore the
random nature of disc-halo interactions.  Ultimately, improvements
in observations together with a more complete survey of models via
simulations should allow us to fully exploit bars in discs as a means
of testing and constraining theories of structure formation.

\section*{Acknowledgements}
{LMW and JB are supported by a Discovery Grant with the Natural
  Sciences and Engineering Research Council of Canada. JSB
  acknowledges the assistance of Matthew Chequers and Keir Darling in
  understanding the \textsc{AGAMA} program interface.}




\bibliographystyle{mnras}
\bibliography{bibliography.bib} 

\begin{thebibliography}{}
\makeatletter
\relax
\def\mn@urlcharsother{\let\do\@makeother \do\$\do\&\do\#\do\^\do\_\do\%\do\~}
\def\mn@doi{\begingroup\mn@urlcharsother \@ifnextchar [ {\mn@doi@}
  {\mn@doi@[]}}
\def\mn@doi@[#1]#2{\def\@tempa{#1}\ifx\@tempa\@empty \href
  {http://dx.doi.org/#2} {doi:#2}\else \href {http://dx.doi.org/#2} {#1}\fi
  \endgroup}
\def\mn@eprint#1#2{\mn@eprint@#1:#2::\@nil}
\def\mn@eprint@arXiv#1{\href {http://arxiv.org/abs/#1} {{\tt arXiv:#1}}}
\def\mn@eprint@dblp#1{\href {http://dblp.uni-trier.de/rec/bibtex/#1.xml}
  {dblp:#1}}
\def\mn@eprint@#1:#2:#3:#4\@nil{\def\@tempa {#1}\def\@tempb {#2}\def\@tempc
  {#3}\ifx \@tempc \@empty \let \@tempc \@tempb \let \@tempb \@tempa \fi \ifx
  \@tempb \@empty \def\@tempb {arXiv}\fi \@ifundefined
  {mn@eprint@\@tempb}{\@tempb:\@tempc}{\expandafter \expandafter \csname
  mn@eprint@\@tempb\endcsname \expandafter{\@tempc}}}

\bibitem[\protect\citeauthoryear{{Athanassoula} \& {Sellwood}}{{Athanassoula}
  \& {Sellwood}}{1986}]{AthanassoulaSellwood1986}
{Athanassoula} E.,  {Sellwood} J.~A.,  1986, \mn@doi [\mnras]
  {10.1093/mnras/221.2.213}, \href
  {http://adsabs.harvard.edu/abs/1986MNRAS.221..213A} {221, 213}

\bibitem[\protect\citeauthoryear{{Bauer}, {Widrow}  \& {Erkal}}{{Bauer}
  et~al.}{2018}]{Bauer2018a}
{Bauer} J.~S.,  {Widrow} L.~M.,   {Erkal} D.,  2018, preprint, \href
  {http://adsabs.harvard.edu/abs/2018arXiv180103608B} {} (\mn@eprint {arXiv}
  {1801.03608})

\bibitem[\protect\citeauthoryear{{Berentzen} \& {Shlosman}}{{Berentzen} \&
  {Shlosman}}{2006}]{BerentzenShlosmanStellarDisks}
{Berentzen} I.,  {Shlosman} I.,  2006, \mn@doi [\apj] {10.1086/506016}, \href
  {http://adsabs.harvard.edu/abs/2006ApJ...648..807B} {648, 807}

\bibitem[\protect\citeauthoryear{{Binney} \& {Tremaine}}{{Binney} \&
  {Tremaine}}{2008}]{BT}
{Binney} J.,  {Tremaine} S.,  2008, {Galactic Dynamics: Second Edition}.
Princeton University Press

\bibitem[\protect\citeauthoryear{{Binney}, {Jiang}  \& {Dutta}}{{Binney}
  et~al.}{1998}]{binney1998}
{Binney} J.,  {Jiang} I.-G.,   {Dutta} S.,  1998, \mn@doi [\mnras]
  {10.1046/j.1365-8711.1998.01595.x}, \href
  {http://adsabs.harvard.edu/abs/1998MNRAS.297.1237B} {297, 1237}

\bibitem[\protect\citeauthoryear{{Blumenthal}, {Faber}, {Flores}  \&
  {Primack}}{{Blumenthal} et~al.}{1986}]{blumenthal1986}
{Blumenthal} G.~R.,  {Faber} S.~M.,  {Flores} R.,   {Primack} J.~R.,  1986,
  \mn@doi [\apj] {10.1086/163867}, \href
  {http://adsabs.harvard.edu/abs/1986ApJ...301...27B} {301, 27}

\bibitem[\protect\citeauthoryear{{Camm}}{{Camm}}{1950}]{camm1950}
{Camm} G.~L.,  1950, \mn@doi [\mnras] {10.1093/mnras/110.4.305}, \href
  {http://adsabs.harvard.edu/abs/1950MNRAS.110..305C} {110, 305}

\bibitem[\protect\citeauthoryear{{Christodoulou}, {Shlosman}  \&
  {Tohline}}{{Christodoulou} et~al.}{1995}]{ChristodoulouStability1995}
{Christodoulou} D.~M.,  {Shlosman} I.,   {Tohline} J.~E.,  1995, \mn@doi [\apj]
  {10.1086/175548}, \href {http://adsabs.harvard.edu/abs/1995ApJ...443..563C}
  {443, 563}

\bibitem[\protect\citeauthoryear{{Combes} \& {Sanders}}{{Combes} \&
  {Sanders}}{1981}]{CombesSandersBars1981}
{Combes} F.,  {Sanders} R.~H.,  1981, \aap, \href
  {http://adsabs.harvard.edu/abs/1981A%26A....96..164C} {96, 164}

\bibitem[\protect\citeauthoryear{{DeBuhr}, {Ma}  \& {White}}{{DeBuhr}
  et~al.}{2012}]{DeBuhrStellarDisks}
{DeBuhr} J.,  {Ma} C.-P.,   {White} S.~D.~M.,  2012, \mn@doi [\mnras]
  {10.1111/j.1365-2966.2012.21910.x}, \href
  {http://adsabs.harvard.edu/abs/2012MNRAS.426..983D} {426, 983}

\bibitem[\protect\citeauthoryear{{Debattista} \& {Sellwood}}{{Debattista} \&
  {Sellwood}}{1998}]{debattista1998}
{Debattista} V.~P.,  {Sellwood} J.~A.,  1998, \mn@doi [\apjl] {10.1086/311118},
  \href {http://adsabs.harvard.edu/abs/1998ApJ...493L...5D} {493, L5}

\bibitem[\protect\citeauthoryear{{Debattista} \& {Sellwood}}{{Debattista} \&
  {Sellwood}}{2000}]{debattista2000}
{Debattista} V.~P.,  {Sellwood} J.~A.,  2000, \mn@doi [\apj] {10.1086/317148},
  \href {http://adsabs.harvard.edu/abs/2000ApJ...543..704D} {543, 704}

\bibitem[\protect\citeauthoryear{{Dorman} et~al.,}{{Dorman}
  et~al.}{2013}]{dorman2013}
{Dorman} C.~E.,  et~al., 2013, \mn@doi [\apj] {10.1088/0004-637X/779/2/103},
  \href {http://adsabs.harvard.edu/abs/2013ApJ...779..103D} {779, 103}

\bibitem[\protect\citeauthoryear{{Dubinski}}{{Dubinski}}{1994}]{dubinski1994}
{Dubinski} J.,  1994, \mn@doi [\apj] {10.1086/174512}, \href
  {http://adsabs.harvard.edu/abs/1994ApJ...431..617D} {431, 617}

\bibitem[\protect\citeauthoryear{{Dubinski} \& {Chakrabarty}}{{Dubinski} \&
  {Chakrabarty}}{2009}]{dubinski2009}
{Dubinski} J.,  {Chakrabarty} D.,  2009, \mn@doi [\apj]
  {10.1088/0004-637X/703/2/2068}, \href
  {http://adsabs.harvard.edu/abs/2009ApJ...703.2068D} {703, 2068}

\bibitem[\protect\citeauthoryear{{Dubinski} \& {Kuijken}}{{Dubinski} \&
  {Kuijken}}{1995a}]{dubinski1995}
{Dubinski} J.,  {Kuijken} K.,  1995a, \mn@doi [\apj] {10.1086/175456}, \href
  {http://adsabs.harvard.edu/abs/1995ApJ...442..492D} {442, 492}

\bibitem[\protect\citeauthoryear{{Dubinski} \& {Kuijken}}{{Dubinski} \&
  {Kuijken}}{1995b}]{DubinskiKuijkenRigidDisks}
{Dubinski} J.,  {Kuijken} K.,  1995b, \mn@doi [\apj] {10.1086/175456}, \href
  {http://adsabs.harvard.edu/abs/1995ApJ...442..492D} {442, 492}

\bibitem[\protect\citeauthoryear{{Dubinski}, {Berentzen}  \&
  {Shlosman}}{{Dubinski} et~al.}{2009}]{dbs2009}
{Dubinski} J.,  {Berentzen} I.,   {Shlosman} I.,  2009, \mn@doi [\apj]
  {10.1088/0004-637X/697/1/293}, \href
  {http://adsabs.harvard.edu/abs/2009ApJ...697..293D} {697, 293}

\bibitem[\protect\citeauthoryear{{Efstathiou}, {Lake}  \&
  {Negroponte}}{{Efstathiou} et~al.}{1982}]{EfstathiouShotNoise}
{Efstathiou} G.,  {Lake} G.,   {Negroponte} J.,  1982, \mn@doi [\mnras]
  {10.1093/mnras/199.4.1069}, \href
  {http://adsabs.harvard.edu/abs/1982MNRAS.199.1069E} {199, 1069}

\bibitem[\protect\citeauthoryear{{Gauthier}, {Dubinski}  \&
  {Widrow}}{{Gauthier} et~al.}{2006}]{gauthier2006}
{Gauthier} J.-R.,  {Dubinski} J.,   {Widrow} L.~M.,  2006, \mn@doi [\apj]
  {10.1086/508860}, \href {http://adsabs.harvard.edu/abs/2006ApJ...653.1180G}
  {653, 1180}

\bibitem[\protect\citeauthoryear{{Goldreich} \& {Tremaine}}{{Goldreich} \&
  {Tremaine}}{1978}]{GoldreichTremaine1978}
{Goldreich} P.,  {Tremaine} S.,  1978, \mn@doi [\apj] {10.1086/156203}, \href
  {http://adsabs.harvard.edu/abs/1978ApJ...222..850G} {222, 850}

\bibitem[\protect\citeauthoryear{{Goldreich} \& {Tremaine}}{{Goldreich} \&
  {Tremaine}}{1979}]{GoldreichTremaine1979}
{Goldreich} P.,  {Tremaine} S.,  1979, \mn@doi [\apj] {10.1086/157448}, \href
  {http://adsabs.harvard.edu/abs/1979ApJ...233..857G} {233, 857}

\bibitem[\protect\citeauthoryear{{Hahn} \& {Abel}}{{Hahn} \&
  {Abel}}{2013}]{music}
{Hahn} O.,  {Abel} T.,  2013, {MUSIC: MUlti-Scale Initial Conditions},
  Astrophysics Source Code Library (\mn@eprint {ascl} {1311.011})

\bibitem[\protect\citeauthoryear{{Hartmann}, {Debattista}, {Cole}, {Valluri},
  {Widrow}  \& {Shen}}{{Hartmann} et~al.}{2014}]{hartmann2014}
{Hartmann} M.,  {Debattista} V.~P.,  {Cole} D.~R.,  {Valluri} M.,  {Widrow}
  L.~M.,   {Shen} J.,  2014, \mn@doi [\mnras] {10.1093/mnras/stu627}, \href
  {http://adsabs.harvard.edu/abs/2014MNRAS.441.1243H} {441, 1243}

\bibitem[\protect\citeauthoryear{{Katz}, {Quinn}, {Bertschinger}  \&
  {Gelb}}{{Katz} et~al.}{1994}]{KatzQuasarZoom}
{Katz} N.,  {Quinn} T.,  {Bertschinger} E.,   {Gelb} J.~M.,  1994, \mn@doi
  [\mnras] {10.1093/mnras/270.1.L71}, \href
  {http://adsabs.harvard.edu/abs/1994MNRAS.270L..71K} {270, L71}

\bibitem[\protect\citeauthoryear{{Kazantzidis}, {Bullock}, {Zentner},
  {Kravtsov}  \& {Moustakas}}{{Kazantzidis} et~al.}{2008}]{kazantzidis2008}
{Kazantzidis} S.,  {Bullock} J.~S.,  {Zentner} A.~R.,  {Kravtsov} A.~V.,
  {Moustakas} L.~A.,  2008, \mn@doi [\apj] {10.1086/591958}, \href
  {http://adsabs.harvard.edu/abs/2008ApJ...688..254K} {688, 254}

\bibitem[\protect\citeauthoryear{{Klypin}, {Valenzuela}, {Col{\'{\i}}n}  \&
  {Quinn}}{{Klypin} et~al.}{2009}]{Klypin2009}
{Klypin} A.,  {Valenzuela} O.,  {Col{\'{\i}}n} P.,   {Quinn} T.,  2009, \mn@doi
  [\mnras] {10.1111/j.1365-2966.2009.15187.x}, \href
  {http://adsabs.harvard.edu/abs/2009MNRAS.398.1027K} {398, 1027}

\bibitem[\protect\citeauthoryear{{Kuijken} \& {Dubinski}}{{Kuijken} \&
  {Dubinski}}{1995}]{GalactICS1995}
{Kuijken} K.,  {Dubinski} J.,  1995, \mn@doi [\mnras]
  {10.1093/mnras/277.4.1341}, \href
  {http://adsabs.harvard.edu/abs/1995MNRAS.277.1341K} {277, 1341}

\bibitem[\protect\citeauthoryear{{Kuijken} \& {Gilmore}}{{Kuijken} \&
  {Gilmore}}{1989}]{KGGalactICSReference}
{Kuijken} K.,  {Gilmore} G.,  1989, \mn@doi [\mnras] {10.1093/mnras/239.2.571},
  \href {http://adsabs.harvard.edu/abs/1989MNRAS.239..571K} {239, 571}

\bibitem[\protect\citeauthoryear{{Martin} et~al.,}{{Martin}
  et~al.}{2014}]{martin2014}
{Martin} N.~F.,  et~al., 2014, \mn@doi [\apj] {10.1088/0004-637X/787/1/19},
  \href {http://adsabs.harvard.edu/abs/2014ApJ...787...19M} {787, 19}

\bibitem[\protect\citeauthoryear{{Martinez-Valpuesta} \&
  {Shlosman}}{{Martinez-Valpuesta} \& {Shlosman}}{2004}]{VP2004}
{Martinez-Valpuesta} I.,  {Shlosman} I.,  2004, \mn@doi [\apjl]
  {10.1086/424876}, \href {http://adsabs.harvard.edu/abs/2004ApJ...613L..29M}
  {613, L29}

\bibitem[\protect\citeauthoryear{{Masters} et~al.,}{{Masters}
  et~al.}{2010}]{masters2010}
{Masters} K.~L.,  et~al., 2010, \mn@doi [\mnras]
  {10.1111/j.1365-2966.2010.16503.x}, \href
  {http://adsabs.harvard.edu/abs/2010MNRAS.405..783M} {405, 783}

\bibitem[\protect\citeauthoryear{{McCarthy}, {Font}, {Crain}, {Deason},
  {Schaye}  \& {Theuns}}{{McCarthy} et~al.}{2012}]{mccarthy2012}
{McCarthy} I.~G.,  {Font} A.~S.,  {Crain} R.~A.,  {Deason} A.~J.,  {Schaye} J.,
    {Theuns} T.,  2012, \mn@doi [\mnras] {10.1111/j.1365-2966.2011.20189.x},
  \href {http://adsabs.harvard.edu/abs/2012MNRAS.420.2245M} {420, 2245}

\bibitem[\protect\citeauthoryear{{Navarro}, {Frenk}  \& {White}}{{Navarro}
  et~al.}{1994}]{NavarroWhiteZoom}
{Navarro} J.~F.,  {Frenk} C.~S.,   {White} S.~D.~M.,  1994, \mn@doi [\mnras]
  {10.1093/mnras/267.1.L1}, \href
  {http://adsabs.harvard.edu/abs/1994MNRAS.267L...1N} {267, L1}

\bibitem[\protect\citeauthoryear{{Navarro}, {Frenk}  \& {White}}{{Navarro}
  et~al.}{1997}]{NFW}
{Navarro} J.~F.,  {Frenk} C.~S.,   {White} S.~D.~M.,  1997, \mn@doi [\apj]
  {10.1086/304888}, \href {http://adsabs.harvard.edu/abs/1997ApJ...490..493N}
  {490, 493}

\bibitem[\protect\citeauthoryear{{Newberg} et~al.,}{{Newberg}
  et~al.}{2002}]{newberg_2002}
{Newberg} H.~J.,  et~al., 2002, \mn@doi [\apj] {10.1086/338983}, \href
  {http://adsabs.harvard.edu/abs/2002ApJ...569..245N} {569, 245}

\bibitem[\protect\citeauthoryear{{O{\~n}orbe}, {Garrison-Kimmel}, {Maller},
  {Bullock}, {Rocha}  \& {Hahn}}{{O{\~n}orbe} et~al.}{2014}]{onorbe_etal_2014}
{O{\~n}orbe} J.,  {Garrison-Kimmel} S.,  {Maller} A.~H.,  {Bullock} J.~S.,
  {Rocha} M.,   {Hahn} O.,  2014, \mn@doi [\mnras] {10.1093/mnras/stt2020},
  \href {http://adsabs.harvard.edu/abs/2014MNRAS.437.1894O} {437, 1894}

\bibitem[\protect\citeauthoryear{{Ostriker} \& {Peebles}}{{Ostriker} \&
  {Peebles}}{1973}]{PeeblesOstriker1973}
{Ostriker} J.~P.,  {Peebles} P.~J.~E.,  1973, \mn@doi [\apj] {10.1086/152513},
  \href {http://adsabs.harvard.edu/abs/1973ApJ...186..467O} {186, 467}

\bibitem[\protect\citeauthoryear{{Planck Collaboration} et~al.,}{{Planck
  Collaboration} et~al.}{2014}]{planck_2014}
{Planck Collaboration} et~al., 2014, \mn@doi [\aap]
  {10.1051/0004-6361/201321591}, \href
  {http://adsabs.harvard.edu/abs/2014A26A...571A..16P} {571, A16}

\bibitem[\protect\citeauthoryear{{Power}, {Navarro}, {Jenkins}, {Frenk},
  {White}, {Springel}, {Stadel}  \& {Quinn}}{{Power}
  et~al.}{2003}]{power_et_al_2003}
{Power} C.,  {Navarro} J.~F.,  {Jenkins} A.,  {Frenk} C.~S.,  {White} S.~D.~M.,
   {Springel} V.,  {Stadel} J.,   {Quinn} T.,  2003, \mn@doi [\mnras]
  {10.1046/j.1365-8711.2003.05925.x}, \href
  {http://adsabs.harvard.edu/abs/2003MNRAS.338...14P} {338, 14}

\bibitem[\protect\citeauthoryear{{Price-Whelan}, {Johnston}, {Sheffield},
  {Laporte}  \& {Sesar}}{{Price-Whelan} et~al.}{2015}]{pricewhelan2015}
{Price-Whelan} A.~M.,  {Johnston} K.~V.,  {Sheffield} A.~A.,  {Laporte}
  C.~F.~P.,   {Sesar} B.,  2015, \mn@doi [\mnras] {10.1093/mnras/stv1324},
  \href {http://adsabs.harvard.edu/abs/2015MNRAS.452..676P} {452, 676}

\bibitem[\protect\citeauthoryear{{Purcell}, {Bullock}  \&
  {Kazantzidis}}{{Purcell} et~al.}{2010}]{purcell2010}
{Purcell} C.~W.,  {Bullock} J.~S.,   {Kazantzidis} S.,  2010, \mn@doi [\mnras]
  {10.1111/j.1365-2966.2010.16429.x}, \href
  {http://adsabs.harvard.edu/abs/2010MNRAS.404.1711P} {404, 1711}

\bibitem[\protect\citeauthoryear{{Purcell}, {Bullock}, {Tollerud}, {Rocha}  \&
  {Chakrabarti}}{{Purcell} et~al.}{2011}]{purcell2011}
{Purcell} C.~W.,  {Bullock} J.~S.,  {Tollerud} E.~J.,  {Rocha} M.,
  {Chakrabarti} S.,  2011, \mn@doi [\nat] {10.1038/nature10417}, \href
  {http://adsabs.harvard.edu/abs/2011Natur.477..301P} {477, 301}

\bibitem[\protect\citeauthoryear{{Read}}{{Read}}{2014}]{read2014}
{Read} J.~I.,  2014, \mn@doi [Journal of Physics G Nuclear Physics]
  {10.1088/0954-3899/41/6/063101}, \href
  {http://adsabs.harvard.edu/abs/2014JPhG...41f3101R} {41, 063101}

\bibitem[\protect\citeauthoryear{{Rocha-Pinto}, {Majewski}, {Skrutskie}  \&
  {Crane}}{{Rocha-Pinto} et~al.}{2003}]{rochapinto2003}
{Rocha-Pinto} H.~J.,  {Majewski} S.~R.,  {Skrutskie} M.~F.,   {Crane} J.~D.,
  2003, \mn@doi [\apjl] {10.1086/378668}, \href
  {http://adsabs.harvard.edu/abs/2003ApJ...594L.115R} {594, L115}

\bibitem[\protect\citeauthoryear{{Ryden} \& {Gunn}}{{Ryden} \&
  {Gunn}}{1987}]{ryden1987}
{Ryden} B.~S.,  {Gunn} J.~E.,  1987, \mn@doi [\apj] {10.1086/165349}, \href
  {http://adsabs.harvard.edu/abs/1987ApJ...318...15R} {318, 15}

\bibitem[\protect\citeauthoryear{{Schaye} et~al.,}{{Schaye}
  et~al.}{2015}]{Eagle}
{Schaye} J.,  et~al., 2015, \mn@doi [\mnras] {10.1093/mnras/stu2058}, \href
  {http://adsabs.harvard.edu/abs/2015MNRAS.446..521S} {446, 521}

\bibitem[\protect\citeauthoryear{{Sellwood}}{{Sellwood}}{1981}]{Sellwood1981}
{Sellwood} J.~A.,  1981, \aap, \href
  {http://adsabs.harvard.edu/abs/1981A%26A....99..362S} {99, 362}

\bibitem[\protect\citeauthoryear{{Sellwood}}{{Sellwood}}{2013}]{Sellwood2013}
{Sellwood} J.~A.,  2013, {Dynamics of Disks and Warps}.
p.~923, \mn@doi{10.1007/978-94-007-5612-0_18}

\bibitem[\protect\citeauthoryear{{Sellwood} \& {Wilkinson}}{{Sellwood} \&
  {Wilkinson}}{1993}]{sellwood1993}
{Sellwood} J.~A.,  {Wilkinson} A.,  1993, \mn@doi [Reports on Progress in
  Physics] {10.1088/0034-4885/56/2/001}, \href
  {http://adsabs.harvard.edu/abs/1993RPPh...56..173S} {56, 173}

\bibitem[\protect\citeauthoryear{{Sharma}, {Johnston}, {Majewski}, {Mu{\~n}oz},
  {Carlberg}  \& {Bullock}}{{Sharma} et~al.}{2010}]{sharma2010}
{Sharma} S.,  {Johnston} K.~V.,  {Majewski} S.~R.,  {Mu{\~n}oz} R.~R.,
  {Carlberg} J.~K.,   {Bullock} J.,  2010, \mn@doi [\apj]
  {10.1088/0004-637X/722/1/750}, \href
  {http://adsabs.harvard.edu/abs/2010ApJ...722..750S} {722, 750}

\bibitem[\protect\citeauthoryear{{Sheffield}, {Price-Whelan}, {Tzanidakis},
  {Johnston}, {Laporte}  \& {Sesar}}{{Sheffield} et~al.}{2018}]{sheffield2018}
{Sheffield} A.~A.,  {Price-Whelan} A.~M.,  {Tzanidakis} A.,  {Johnston} K.~V.,
  {Laporte} C.~F.~P.,   {Sesar} B.,  2018, \mn@doi [\apj]
  {10.3847/1538-4357/aaa4b6}, \href
  {http://adsabs.harvard.edu/abs/2018ApJ...854...47S} {854, 47}

\bibitem[\protect\citeauthoryear{{Simmons} et~al.,}{{Simmons}
  et~al.}{2014}]{simmons2014}
{Simmons} B.~D.,  et~al., 2014, \mn@doi [\mnras] {10.1093/mnras/stu1817}, \href
  {http://adsabs.harvard.edu/abs/2014MNRAS.445.3466S} {445, 3466}

\bibitem[\protect\citeauthoryear{{Spitzer}}{{Spitzer}}{1942}]{spitzer1942}
{Spitzer} Jr. L.,  1942, \mn@doi [\apj] {10.1086/144407}, \href
  {http://adsabs.harvard.edu/abs/1942ApJ....95..329S} {95, 329}

\bibitem[\protect\citeauthoryear{{Springel}}{{Springel}}{2005}]{GadgetCodePaper}
{Springel} V.,  2005, \mn@doi [\mnras] {10.1111/j.1365-2966.2005.09655.x},
  \href {http://adsabs.harvard.edu/abs/2005MNRAS.364.1105S} {364, 1105}

\bibitem[\protect\citeauthoryear{{Tissera}, {Scannapieco}, {Beers}  \&
  {Carollo}}{{Tissera} et~al.}{2013}]{tissera2013}
{Tissera} P.~B.,  {Scannapieco} C.,  {Beers} T.~C.,   {Carollo} D.,  2013,
  \mn@doi [\mnras] {10.1093/mnras/stt691}, \href
  {http://adsabs.harvard.edu/abs/2013MNRAS.432.3391T} {432, 3391}

\bibitem[\protect\citeauthoryear{{Toomre}}{{Toomre}}{1964}]{ToomreParameter}
{Toomre} A.,  1964, \mn@doi [\apj] {10.1086/147861}, \href
  {http://adsabs.harvard.edu/abs/1964ApJ...139.1217T} {139, 1217}

\bibitem[\protect\citeauthoryear{{Vasiliev}}{{Vasiliev}}{2018}]{agama}
{Vasiliev} E.,  2018, preprint, \href
  {http://adsabs.harvard.edu/abs/2018arXiv180208239V} {} (\mn@eprint {arXiv}
  {1802.08239})

\bibitem[\protect\citeauthoryear{{Vogelsberger}, {Genel}, {Sijacki}, {Torrey},
  {Springel}  \& {Hernquist}}{{Vogelsberger} et~al.}{2013}]{IllustrisFeedback}
{Vogelsberger} M.,  {Genel} S.,  {Sijacki} D.,  {Torrey} P.,  {Springel} V.,
  {Hernquist} L.,  2013, \mn@doi [\mnras] {10.1093/mnras/stt1789}, \href
  {http://adsabs.harvard.edu/abs/2013MNRAS.436.3031V} {436, 3031}

\bibitem[\protect\citeauthoryear{{Widrow}, {Pym}  \& {Dubinski}}{{Widrow}
  et~al.}{2008}]{WPDGalactICSReference}
{Widrow} L.~M.,  {Pym} B.,   {Dubinski} J.,  2008, \mn@doi [\apj]
  {10.1086/587636}, \href {http://adsabs.harvard.edu/abs/2008ApJ...679.1239W}
  {679, 1239}

\bibitem[\protect\citeauthoryear{{Yanny} et~al.,}{{Yanny}
  et~al.}{2000}]{yanny2000}
{Yanny} B.,  et~al., 2000, \mn@doi [\apj] {10.1086/309386}, \href
  {http://adsabs.harvard.edu/abs/2000ApJ...540..825Y} {540, 825}

\bibitem[\protect\citeauthoryear{{Yurin} \& {Springel}}{{Yurin} \&
  {Springel}}{2015}]{YurinSpringelStellarDisks}
{Yurin} D.,  {Springel} V.,  2015, \mn@doi [\mnras] {10.1093/mnras/stv1454},
  \href {http://adsabs.harvard.edu/abs/2015MNRAS.452.2367Y} {452, 2367}

\bibitem[\protect\citeauthoryear{{Zang} \& {Hohl}}{{Zang} \&
  {Hohl}}{1978}]{ZangHohlBars1978}
{Zang} T.~A.,  {Hohl} F.,  1978, \mn@doi [\apj] {10.1086/156636}, \href
  {http://adsabs.harvard.edu/abs/1978ApJ...226..521Z} {226, 521}

\makeatother
\end{thebibliography}


\bsp	
\label{lastpage}

\end{document}